\documentclass[aps,prd,reprint,superscriptaddress,unsortedaddress]{revtex4-1} 
\usepackage{graphicx}
\usepackage{amssymb,amsmath,amsfonts,ifthen}
\usepackage{multirow}
\usepackage{array}
\usepackage{dcolumn}
\newcolumntype{!}{D{!}{\,\pm\,}{3,3}}
\newcolumntype{.}{D}
\DeclareMathSymbol{\minus}{\mathord}{operators}{"2D}
\providecommand{\oforder}{\ensuremath{\sim\!\!}}
\providecommand{\e}[1]{\ensuremath{\!\times\!10^{#1}}}
\providecommand{\cov}{\ensuremath{\text{Cov}}}
\providecommand{\var}{\ensuremath{\text{Var}}}
\usepackage{enumerate}

\usepackage[final, ulem={normalem,normalbf}]{changes} 

\definechangesauthor[Michele Armano]{MA}{red}

\usepackage{units}

\begin{document}

\listofchanges


\preprint{DRAFT}

\title{Time domain maximum likelihood parameter estimation \\ in LISA Pathfinder Data Analysis}

\author{G.~Congedo}
\email[Electronic address: ]{congedo@science.unitn.it}

\author{L.~Ferraioli}
\author{M.~Hueller}
\author{F.~De Marchi}
\author{S.~Vitale}
\affiliation{Dipartimento di Fisica, Universit\`a di Trento and INFN,
             Gruppo Collegato di Trento, \\
             38123 Povo, Trento, Italy}

\author{M.~Armano}
\affiliation{European Space Agency (ESA), SRE-OD ESAC \\
             28692 Camino bajo del Castillo, Villanueva de la Ca\~{n}ada, Madrid}

\author{M.~Hewitson}
\affiliation{Albert-Einstein-Institut, Max-Planck-Institut f\"ur
             Gravitationsphysik und Universit\"at Hannover, \\
             30167 Hannover, Germany}
             
\author{M.~Nofrarias}
\affiliation{Institut de Ci\`encies de l'Espai, (CSIC-IEEC), \\
             Facultat de Ci\`encies, Campus UAB, Torre C-5, \\
             08193 Bellaterra, Barcelona, Spain}

\date{\today}

\begin{abstract}

LISA is the upcoming space-based Gravitational Wave telescope.  LISA Pathfinder, to
be launched in the coming years, will prove and verify the detection principle
of the fundamental Doppler link of LISA on a flight hardware identical in design to that of LISA.
LISA Pathfinder will collect a picture of all noise disturbances possibly affecting LISA, achieving the unprecedented pureness of geodesic motion necessary
for the detection of gravitational waves. The first steps of both missions will crucially depend on a very precise calibration
of the key system parameters. Moreover, robust parameters estimation is of fundamental importance in the correct assessment of the residual force noise,
an essential part of the data processing for LISA. In this paper we present a maximum likelihood parameter estimation technique in time domain being devised for this calibration
and show its proficiency on simulated data and validation through Monte Carlo realizations of independent noise runs. We discuss its robustness to
non-standard scenarios possibly arising during the real-life mission, as well
as its independence to the initial guess and non-gaussianities. Furthermore, we apply the same technique to data produced in mission-like fashion during operational exercises with a realistic
simulator provided by ESA.

\end{abstract}


\maketitle


\section{Introduction}

LISA \cite{bender1998,danzmann2011} is the proposed space-based Gravitational Waves
(GWs) observatory planned to fly by the next decade. It is based on three SpaceCrafts (SCs) --- each hosting and protecting two
Test Masses (TMs) in nominal free fall --- flying in a 5 Million
km sided triangular formation around the Sun at 1 AU. A total of 6 TMs, whose displacements are detected by a laser-interferometric
technique, constitute 6 Doppler links, two per LISA arm, tracking
the local curvature variations around the Sun and being sensitive to the small fluctuations
induced by GW signals in the $\unit[0.1-100]{mHz}$ band.

One (any) arm of LISA is virtually shrunk \cite{vitale2009} to $\unit[38]{cm}$ and implemented in the
LISA Pathfinder (LPF) mission \cite{vitale2002,armano2009,antonucci2011a}. LPF will effectively
measure the differential force noise that pollutes the sensitivity of LISA below $\unit[3\e{\minus14}]{m\,s^{\minus2}\,Hz^{\minus1/2}}$
around $\unit[1]{mHz}$ --- the minimum performance level for LISA to carry on its science program in astrophysics.

The observational horizon of LISA will include thousands of GW sources. Among all, the highest Signal-to-Noise sources will be surely the Super-Massive Black Holes (SMBHs). However, there are sources that are at the limit of the LISA sensitivity for which an accurate assessment of the instrumental noise is mandatory. The population of the Extreme Mass Ratio Inspirals (EMRIs) \cite{amaro2007} is the most important example: they are a valuable instrument to test general relativity and curvature in the strong gravity regime. Different EMRI search methods have been developed. After having subtracted the highest signals (SMBHs and calibration binaries), in order to extract the EMRI signatures, all methods strictly have to deal with the instrumental noise level, for which the LPF mission has a crucial role. In fact, a systematic error in the reconstructed noise shape would dramatically affect the identification of such sources. The methods described in this paper allows for a solution of this problem.

The main payload on-board LPF, the LISA Technology Package (LTP) \cite{armano2009},
will thus be used in an extensive characterization campaign by
measuring all force disturbances and systematics. To the purpose, a precise
calibration of the key system parameters must be performed before any assessment
of the final level of differential force noise can be made. The full process
is iterative: the quality of free fall achieved at a given stage of the mission depends
on the results of the previous experiments. By proceeding in the direction of increasing
precision, the observed noise will be fully explained.

In LPF as a physical system the relationship between sensed displacements and applied station-keeping forces plays a crucial role. Hence, the effect of such forces must be taken into account and subtracted from the data, in order to provide a successful estimate of the external residual force noise. To this end and to invert the system dynamics the calibration of all key system parameters is required, a problem we address and solve by maximum likelihood parameter estimation in time domain. Preliminary work was presented in \cite{congedo2010}, we hereby extend the method, present it in a more robust fashion and apply it to simulated datasets as well as to more realistic simulation data released by ESA.

The paper is structured as follows. In Section \ref{sect:dynamics} we provide a general description of a multi-controlled dynamical system and show the procedure to obtain the external residual out-of-loop force noise. Then, we apply the formalism to the LPF mission and provide a model for the dynamics along the two most relevant degrees of freedom.
In Section \ref{sect:maximum_likelihood} we demonstrate how the method is capable of correctly identifying all parameters and handling the problem of degeneracy by collecting the information from different experiments aimed at exciting different degrees of freedom of the system. In Section \ref{sect:data_analysis} we report on the results of our investigations. More in details, in Section \ref{sect:data_generation} we discuss the data production, in \ref{sect:whitening_filters} the whitening filters and finally in \ref{sect:param_est} the parameter estimation. Section \ref{sect:monte_carlo} presents the Monte Carlo validation of the method; \ref{sect:robust2guess} and \ref{sect:robust2glitches} describe the proficiency on applying the same technique to non-standard scenarios, corresponding to a poorly calibrated (or strongly under-performing) system (robustness to the initial guess) and a readout affected by glitches (robustness to non-gaussianities). Section \ref{sect:stoc} provides an example of analysis of data produced by the ESA LPF science simulator, more realistic but treated in part as a black-box. Finally, in Section \ref{sect:force_noise} we discuss the overall impact of the method to the estimation of the residual force noise.


\section{Dynamics and system identification} \label{sect:dynamics}

The LISA link is ideally composed of two TMs working as mirrors and whose relative displacement is tracked by a laser interferometer. While controlled along the other degrees of freedom, the TMs are left in nominal free-fall along the optically sensed axis. In LISA there is actually no direct measurement of the differential TM displacement, but a combination of local measurements (TM to local optical bench) and the one between two far apart SCs.

LPF holds two main conceptual differences with respect to LISA: the differential
motion is directly detected by a laser interferometer and the second TM is
controlled along the sensitive axis. Indeed, in the main science mode, LPF is a
controlled dynamical system where the reference TM is in free fall along the sensitive
axis and electrostatically suspended along the other degrees of freedom.
Interferometric readouts are used by the controller to compute specific commands sent to the actuators to drive the SC and the second TM to
follow the reference TM. The control of the separation of the SC to the reference TM is called drag-free loop; the control of the separation of
the second TM and the reference TM is called electrostatic suspension loop. In this way, LPF is actually a dynamical system with coupled control loops.

The full equation of motion expressed in the interferometer sensing coordinates ($\mathbf{o}$) can be written as (see Appendix \ref{app:dyn} for details)
\begin{equation}
\mathbf{\Delta}\cdot\mathbf{o} = \mathbf{D}\cdot\mathbf{S}^{\minus1}\cdot\mathbf{o}_\text{n}+\mathbf{f}+\mathbf{C}\cdot\mathbf{T}\cdot\mathbf{o}_\text{i}~, \label{eq:eom_full}
\end{equation}
where $\mathbf{D}$ contains the derivatives, $\mathbf{C}$ is the control matrix, $\mathbf{T}$ maps the delays and $\mathbf{S}$ the sensing strategy. $\mathbf{o}_\text{i}$ are setpoint injections in the
interferometer channels and $\mathbf{o}_\text{n}$ is the readout noise vector. $\mathbf{f}$ represent the external forces. We have also defined the second-order differentiation operator
\begin{equation}
\mathbf{\Delta} = \mathbf{D}\cdot\mathbf{S}^{\minus1}+\mathbf{C}~. \label{eq:diff_op}
\end{equation}
Two transfer function matrices can be naturally identified
\begin{align}
\mathbf{H}_{\mathbf{o}_\text{i}\rightarrow\mathbf{o}} & = \mathbf{\Delta}^{\minus1}\cdot\mathbf{C}\cdot\mathbf{T}~, \label{eq:ifo2ifo} \\
\mathbf{H}_{\mathbf{o}\rightarrow\mathbf{f}} & = \mathbf{\Delta}~. \label{eq:ifo2acc}
\end{align}
In particular, the second is of fundamental relevance as it shows that the differentiation operator allows to estimate the out-of-loop external residual force per unit mass. However, such evaluation requires to handle the effect of the controller and calibrate the parameters contained in the matrices $\mathbf{D}$, $\mathbf{S}$ and $\mathbf{C}$. Hence, the first transfer function is used for the system identification or, equivalently, for the estimation of all system parameters in dedicated experiments; the second to estimate the force per unit mass by applying the calibrated operator $\mathbf{\Delta}$ on data with interferometer noise only (all deterministic inputs are set to zero).

Considering the model along the optically-sensed axis described in Eq.\;\eqref{eq:eom_full} and referring to Eq.\;\eqref{eq:eom_1D} for the notation, the computation of the residual force noise requires the calibration of the following minimal set of parameters:
\begin{itemize}
\item $\omega_1^2$ and $\omega_{12}^2$: residual oscillator-like couplings between the SC and the reference TM and between the two TMs, the first one typically $\oforder\unit[1\e{\minus6}]{s^{\minus2}}$;
\item $S_{21}$: the sensing cross-talk between $o_1$ and $o_{12}$ channels, typically $\oforder 1\e{\minus4}$;
\item $A_\mathrm{df}$ and $A_\mathrm{sus}$: actuation gains for application of the forces by the thrusters and the electrostatic suspensions, typically $\oforder 1$;
\item $\Delta t_1$ and $\Delta t_2$: delays in the application of the actuation of the forces by the thrusters and the electrostatic suspensions, typically some fraction of a second.
\end{itemize}

Amongst the series of experiments characterizing the LTP, a few of capital importance will tackle the measurement of the mentioned parameters. We hereby consider two main identification experiments where the
injections are signals modifying the interferometer zero point:
\begin{enumerate}
\item injection into the controller guidance of the $o_1$ channel, namely $o_{\text{i},1}$;
\item injection into the controller guidance of the $o_{12}$ channel, namely $o_{\text{i},12}$.
\end{enumerate}
Clearly the $o_1$ readout of the first experiment is highly
sensitive to $\omega_1^2$, $A_\mathrm{df}$ and $\Delta t_1$; the $o_{12}$
readout is sensitive to almost all parameters, in particular to $S_{21}$, $A_\mathrm{sus}$ and $\Delta t_1$. However, in the second
experiment the $o_1$ readout serves as a sanity check since it is expected to
carry no information (the cross-talk from the differential channel to the first
one is negligible). The combination of the information from
the two available experiments makes the identification of all $7$ parameters feasible.


\section{Maximum likelihood estimation in time domain} \label{sect:maximum_likelihood}

The colored noise shapes expected in the LPF mission force us to develop a rather
general formalism to identify the system parameters.

Let us suppose that our measurement is stored in the time series $(t_i,x_i)$ with
$i=1,...,N_\text{data}$, and the model to fit is $\hat{x}(t_i,\mathbf{p})$,
$\mathbf{p}$ being the vector of all parameters. We can define
the residuals between data and model as $r_i=x_i-\hat{x}(t_i,\mathbf{p})$.
The general recipe is to build the likelihood estimator
\begin{equation}\label{eq:likelihood}
\mathcal{L}(\text{data}\vert\mathbf{p})=\text{const}\; e^{\minus\chi^2(\mathbf{p})/2}~,
\end{equation}
where the argument is the norm of residuals (log-likelihood)
\begin{equation}\label{eq:chi2}
\chi^2(\mathbf{p})=\langle\mathbf{r}(\mathbf{p})\vert\mathbf{r}(\mathbf{p})\rangle~,
\end{equation}
easily identifiable with the usual least square estimator when the noise is uncorrelated and Gaussian
distributed.
The inner product $\langle\cdot\vert\cdot\rangle$ that defines the norm in the preceding equation for discrete values is given by
\begin{equation}\label{eq:inner_dot}
\langle g \vert h \rangle=
\mathbf{g}^\dag \cdot \boldsymbol{\mathcal{C}}_\text{n}^{\minus1} \cdot \mathbf{h}~,
\end{equation}
where $\mathbf{g}$ and $\mathbf{h}$ denote two generic functions evaluated at discrete times and $\boldsymbol{\mathcal{C}}_\text{n}$ is the noise covariance matrix.

Therefore, the parameter estimation task for LPF consists of
either maximizing the likelihood in Eq.\;\eqref{eq:likelihood} or minimizing the
norm (log-likelihood) in Eq.\;\eqref{eq:chi2}.

\subsection{Multi-experiment analysis}

As described in Section \ref{sect:dynamics}, the simplest system identification consists
of at least two experiments (with two interferometer readings each) to be
performed in-flight. Therefore, it is necessary to develop a general estimation method
that includes the information coming from many experiments and measurement channels. This
can also solve the issue of the possible parameter degeneracy by increasing
the total information of the system.

The first intuitive attempt is to fit each experiment (or even each single channel) independently, obtain the various parameter estimates and combine them to get the final result. The underlying philosophy is to accumulate information about the full parameter set. Let us recall the definition of the Fisher information matrix on the best-fit estimates defined as the Hessian of the log-likelihood
\begin{equation}
\mathcal{I}_{kl} = \frac{\partial^2}{\partial p_k \partial p_l} \chi^2(\mathbf{p})~.
\end{equation}
Let us also suppose that $\mathbf{p}_{ij}$ are the parameter estimates, for $i$ and $j$ counting the experiments and the readings per each experiment and $\boldsymbol{\mathcal{I}}_{ij}$ is the relative Fisher information matrix. It can be shown \citep{nofrarias2010} that the combined estimate is an information-weighted average
\begin{equation}\label{eq:comb_estimates}
\mathbf{p}=\boldsymbol{\mathcal{I}}^{\minus1}\cdot\sum_{i=1}^{N_\text{exps}}\sum_{j=1}^{N_\text{chs}}\boldsymbol{\mathcal{I}}_{ij}\cdot\mathbf{p}_{ij}~,
\end{equation}
where the total Fisher information matrix is
\begin{equation}
\boldsymbol{\mathcal{I}}=\sum_{i=1}^{N_\text{exps}}\sum_{j=1}^{N_\text{chs}}\boldsymbol{\mathcal{I}}_{ij}~.
\end{equation}
In terms of a covariance matrix, the preceding formulae are a generalization of the covariance-weighted mean.

Following the principle of Ockham's razor, one usually wants to fit the smallest set of parameters. In Section \ref{sect:dynamics} we concluded our discussion saying that there are some readouts that are more sensitive to some parameters than others. Those others play the role of nuisance parameters for the fit of that readout. Hence it is a good practice to fit only a restricted number of parameters for each readout, those that are more meaningful for it. Clearly, all independent fits provide different estimates to different parameters, yet, some parameters may be shared between experiments. When summing up within Eq.\;\eqref{eq:comb_estimates} we take care of the different matrix sizes by putting zeros where we have no measurement or information.

An alternative method for the multi-experiment parameter estimation is based on building a joint likelihood of all experimental outputs and models for each experiment. Since the typical identification experiment on flight
will not last enough to let the cross correlation become important between
different channels (and surely between different experiments), the hypothesis of
statistical independency is reasonable and the joint likelihood is given by
\begin{equation}\label{eq:joint_likelihood}
\mathcal{L}(\text{data}\vert\mathbf{p})=\prod_{i=1}^{N_\text{exps}}\prod_{j=1}^{N_\text{chs}}\mathcal{L}_{ij}(\text{data}_{\,ij}\vert\mathbf{p})~,
\end{equation}
where $i$ counts the experiments and $j$ the channels per experiments. Analogously,
the joint log-likelihood norm is
\begin{equation}\label{eq:joint_chi2}
\chi^2(\mathbf{p})=\sum_{i=1}^{N_\text{exps}}\sum_{j=1}^{N_\text{chs}}\chi^2_{ij}(\mathbf{p})~.
\end{equation}
Assuming that all channels are sampled at the same rate and last for the same duration, the overall number $\nu$ of degrees of freedom are defined as $\nu=N_\text{exps} \times N_\text{chs} \times N_\text{data} - N_\mathbf{p}$, where $N_\mathbf{p}$ is the dimension of the parameter space.

Our analysis has shown that on simulated data the joint fit and the combination of independent fits provide compatible estimates within the confidence level. However, the inaccuracy of a model for a channel might bias the fit. When this occurs, the information weighed mean of Eq.\;\eqref{eq:comb_estimates} is not robust and it amplifies the bias in the combined estimate. In this case, one can try to remove that estimate and combine the remaining: however, by doing so information and precision would definitely be lost. The joint analysis is more robust to such kind of problems: the poor
information from the badly-fitted model is compensated by the others. We will therefore adopt the joint approach for the rest of the paper.

\subsection{Whitening} \label{sect:whitening}

Let us consider, for simplicity, the case of one experiment with only one reading and assume that stationarity holds true. (In general, one needs to consider also the cross correlation between different channels and experiments.) Eq.\;\eqref{eq:chi2} can be rewritten
in term of the self-inner product of the residual vector. Then, without loss of generality, there exists an orthogonal matrix $\boldsymbol{\mathcal{U}}$ and a diagonal matrix $\mathbf{\Lambda}_\text{n}$ such that
\begin{align}\label{eq:chi2_diag1}
\chi^2(\mathbf{p}) & =\mathbf{r}^\dag(\mathbf{p})\cdot\boldsymbol{\mathcal{C}}_\text{n}^{\minus1}\cdot\mathbf{r}(\mathbf{p}) \nonumber \\
  & =\mathbf{r}^\dag(\mathbf{p})\cdot\left(\boldsymbol{\mathcal{U}}\cdot\mathbf{\Lambda}_\text{n}^{\minus1}\cdot\boldsymbol{\mathcal{U}}^\dag\right)\cdot\mathbf{r}(\mathbf{p}) \nonumber \\
  & =\left(\boldsymbol{\mathcal{U}}^\dag\cdot\mathbf{r}(\mathbf{p})\right)^\dag\cdot\mathbf{\Lambda}_\text{n}^{\minus1}\cdot\left(\boldsymbol{\mathcal{U}}^\dag\cdot\mathbf{r}(\mathbf{p})\right)~.
\end{align}
By diagonalizing the noise covariance matrix, data get decorrelated and this is equivalent to whitening the data in the frequency domain. The likelihood estimation shall now be performed on the transformed residuals, obtained by the application of the operator $\boldsymbol{\mathcal{U}}^\dag$ (the whitening filter), on the residual vector $\mathbf{r}(\mathbf{p})$, along the eigen-direction of the noise.

The same formalism can be applied to the calculation of noise generating functions \cite{ferraioli2010a}.


\subsection{The estimation method}

The parameter estimation templates, to be compared to the experimental data, are calculated from the available matrix $\mathbf{H}(\omega,\mathbf{p})$ of transfer function models. For an experiment having $N$ inputs, contained in the vector $\mathbf{i}$, and $M$ outputs, contained in the vector $\mathbf{o}$, the matrix has size $N \times M$. The modeled ($\hat{\;}$ symbol) system outputs, are given by
\begin{equation}\label{eq:template}
\hat{\mathbf{o}}(t,\mathbf{p})=\mathcal{F}^{\minus1}\left[\mathbf{H}(\omega,\mathbf{p})\cdot\mathbf{i}(\omega)\right](t)~,
\end{equation}
where $\mathcal{F}^{\minus1}[\cdot](t)$ stands for the standard inverse Fourier transform. Here
we assume that all parameter information is contained within the transfer function
matrix and the inputs are not parametric. Indeed, we want to study the system and the injections are usually fully
known.

In order to compute the Fisher information matrix, one also needs the first-order derivatives of the models. This is obtained by the following
\begin{equation}\label{eq:template_der}
\nabla_\mathbf{p} \hat{\mathbf{o}}(t,\mathbf{p})=\mathcal{F}^{\minus1}\left[\nabla_\mathbf{p} \mathbf{H}(\omega,\mathbf{p})\cdot\mathbf{i}(\omega)\right](t)~,
\end{equation}
where $\nabla_{\mathbf{p}}$ is the vector of derivatives with respect to each component of $\mathbf{p}$. This quantity is usually called the model Jacobian. The whitened Jacobian multiplied by its transpose gives the information matrix.

Therefore, it's easy to identify the iteration steps needed to produce the final parameter estimates, in loops of increasing accuracy:
\begin{enumerate}
  \item the whitening filters are estimated from a long noise run;
  \item data are whitened;
  \item the templates are generated from the transfer functions matrix (see Eq.\;\eqref{eq:template});
  \item the templates are whitened;
  \item data are fitted by adjusting the parameters iteratively and generating new whitened templates (step 3 through 4).
\end{enumerate}


\section{Data analysis} \label{sect:data_analysis}

In this section we want to validate the method introduced above by applying it to
mock data. Firstly, we describe the data generation: a long noise run and some injection experiments are simulated. Then, we show the whitening/decorrelation process: here whitening filters are provided for the estimation. Thereafter, we go through the
core step of the parameter estimation where we describe the optimization scheme
and show some results. We prove its consistency with a Monte Carlo simulation and its
robustness to altered boundary conditions: namely the initial guess (corresponding to an under-performing system configuration), and non-gaussianities (the presence of glitches in the readout). Then, we will show the results of the application of the same methodology to data produced by a realistic LPF simulator provided by ESA. Finally, we discuss the relevance of these methods in correctly computing the external residual force noise per unit mass.

Data production and analysis are performed with the LISA Technology Package Data Analysis Toolbox (LTPDA) \cite{ltpda}, an object-oriented extension of MATLAB$^\circledR$ \cite{matlab}.

\subsection{Experiments and data generation} \label{sect:data_generation}

A long noise run lasting 28 hours has been produced by coloring white Gaussian noise with realistic noise shaping filters \cite{ferraioli2010a} and assuming stationarity and normal distribution. During the real mission, this long noise run has two uses: to investigate the noise behavior, and to produce whitening filters for the parameter estimation.

It is worth making some comments on the assumption about the noise stationarity.
For LPF, noise is a stochastic process that we can model as a function
of time and some system parameters, say $n=n\left(t,p(t)\right)$ in the case of a single parameter $p$. A fluctuation $\delta p$ around the
nominal value $p_0$ due to a time-dependency gives --- to first order --- $n \simeq n_0+n'\delta p$, where $n_0=n(t,p_0)$
and $n'=\left.\partial n(t,p)/\partial p\right|_{p_0}$. Then, for a zero-mean process,
the total noise variance is
\begin{equation}\label{eq:noise_variance}
\var[n] \simeq \var[n_0]+\var'[n_0]\delta p+\var[n']\delta p^2~,
\end{equation}
where the linear and quadratic terms come from the covariance between $n_0$ and $n'$
and the variance of $n'$ itself (see Appendix \ref{app:demo} for details). Therefore, if any of the system parameters changes in
time, the noise is likely to become non-stationary. The converse (i.e., non-stationarity implies a variation of the parameters) is
not true, since other effects, independent from those parameters, may still be relevant. For example (Section \ref{sect:robust2glitches}), glitches are a non-stationary noise behavior intrinsic in the readout.

We simulated both experiments described at the end of Section \ref{sect:dynamics} for a total duration of $3$ hours each. We implemented a different and independent noise realization from the previous long noise run, produced deterministic signals following the recipe of Eq.\;\eqref{eq:template} and assumed that the superposition principle of signals and noise holds true in the hypothesis of small motion and in absence of non-linearities.

\begin{figure}[htb]
\includegraphics[width=\columnwidth]{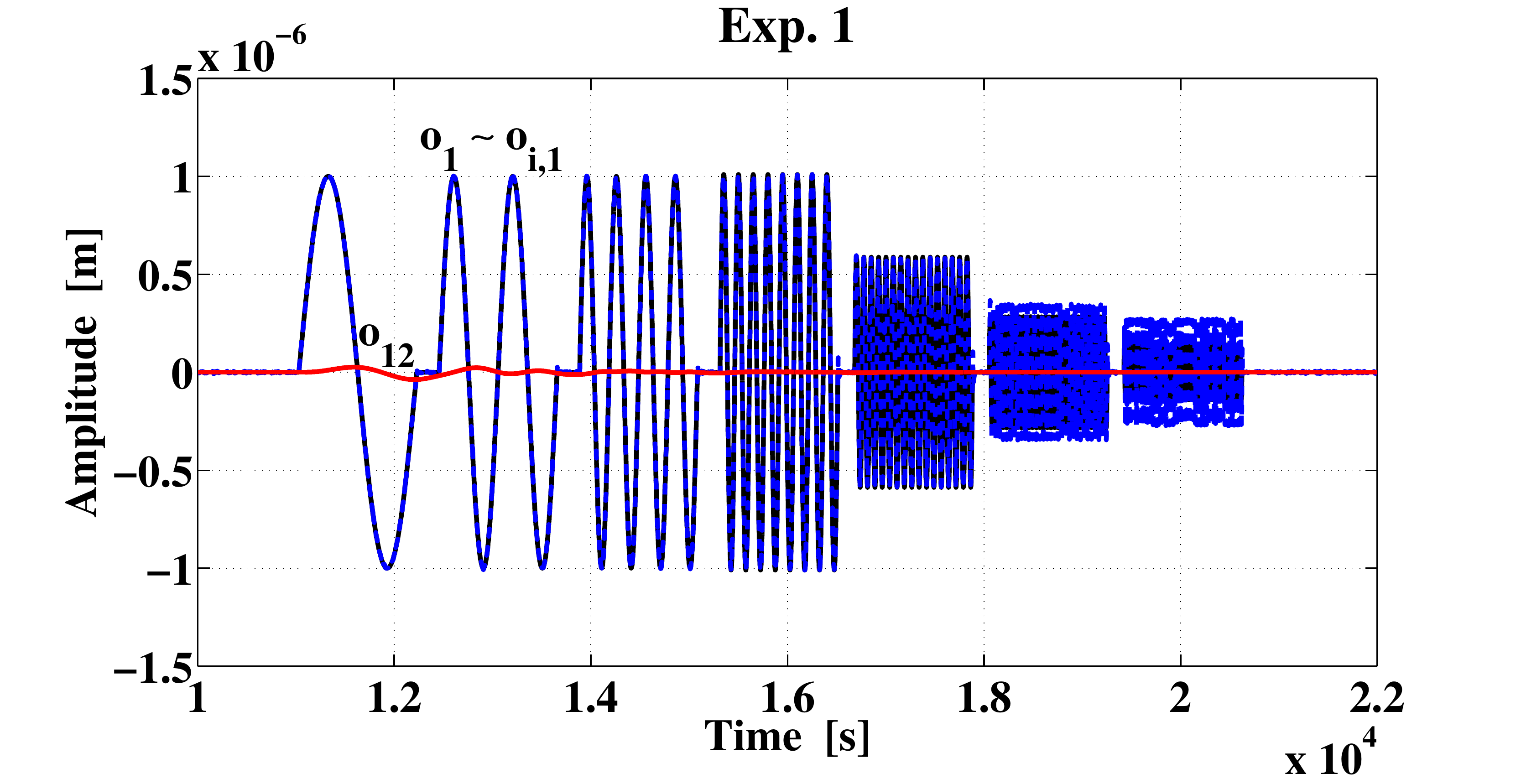} \\
\includegraphics[width=\columnwidth]{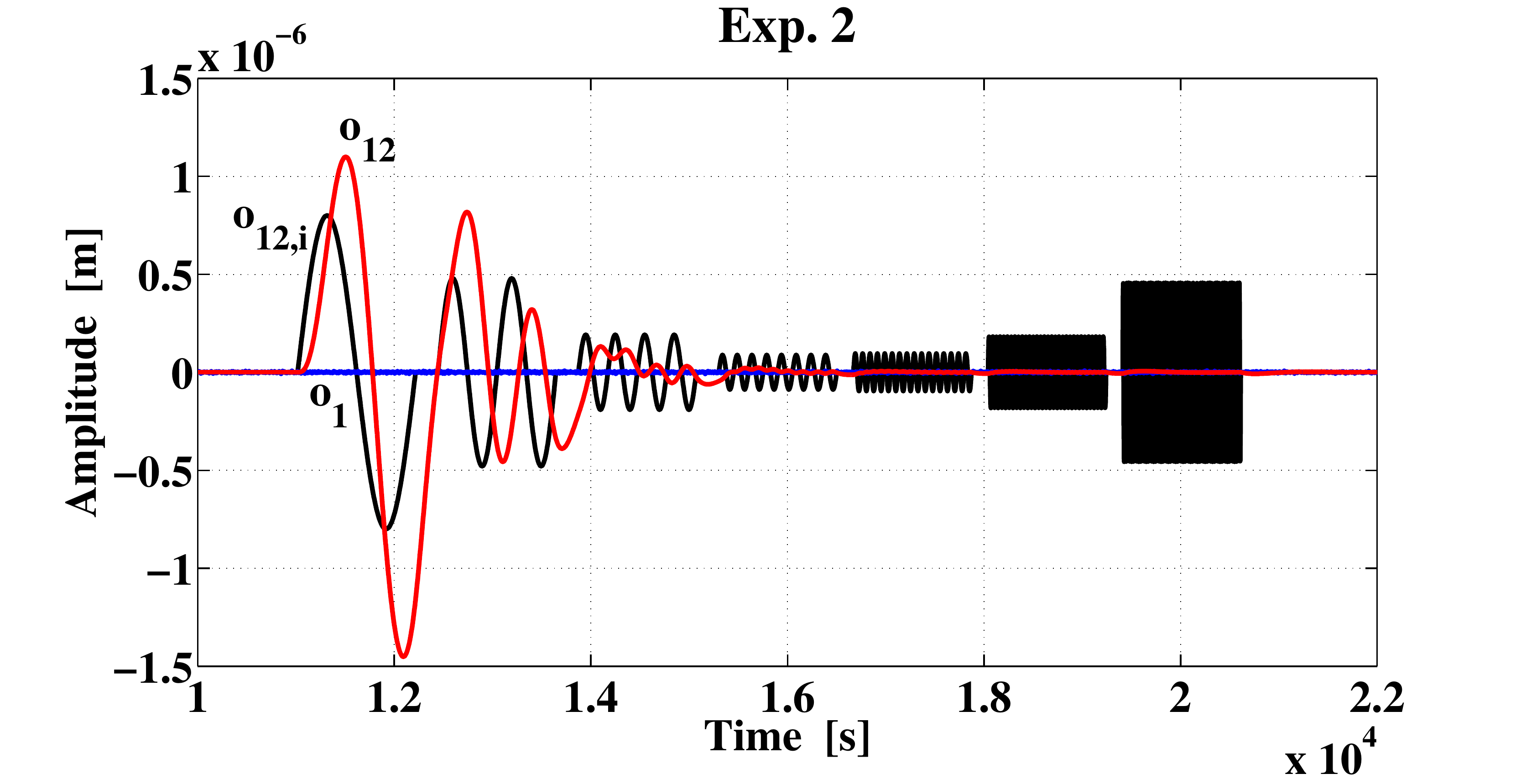}
\caption{\label{fig:data}\footnotesize{Synthetic data generated for Exp.\,1, injection into the first controller guidance $o_{\text{i},1}$, and Exp.\,2, injection into the second controller guidance $o_{\text{i},12}$. We show the response of the system in both channels $o_1$ and $o_{12}$ for the two experiments, expected to not exceed $1\;\mu\text{m}$. We split the data to highlight the part concerning the parameter estimation. In Exp.\,1 the response of the system in $o_1$ (dashed line) is approximately equal to $o_{\text{i},1}$, except at high frequency, and a residual signal in $o_{12}$ is due to the cross-talk. In Exp.\,2 the response of the system in $o_1$ is negligible and in $o_{12}$ is important at low frequency.}}
\end{figure}

An example of the two generated experiments with the respective injections is shown in Fig.\;\ref{fig:data}. A logarithmic sweep of increasing frequency $f=\unit[8.3\e{\minus4},\,1.7\e{\minus3},\,3.3\e{\minus3},\,6.7\e{\minus3},\,1.3\e{\minus2},\,2.7\e{\minus2},\,5.3\e{\minus2}]{Hz}$ (from 1 up to 64 cycles, each single sine stretch lasting $1200\,\unit{s}$), is injected into the system, once in the first guidance $o_{\text{i},1}$, then in the second $o_{\text{i},12}$, to produce the two experiments. The amplitudes are selected not to exceed $1\%$ of the operating range of the instrument and $10\%$ of the maximum allowed forces. In the first experiment the system response in the $o_1$ channel resembles the
original injection $o_{\text{i},1}$, except at high frequency where there is a slight difference. At higher frequencies the system response increases due to the rising of the transfer function, therefore the amplitudes are lower than the corresponding ones at lower frequencies. Instead, the $o_{12}$ channel contains an extra contribution due to the readout cross-talk. In the second experiment the system responds only in the $o_{12}$ channel and the $o_1$ signal content is negligible. Since the system response decays at higher frequencies we must provide injection signals with higher amplitude to see any effect. A phase delay is also present.

In what follows, data have been produced at $\unit[10]{Hz}$ and then (since the highest frequency injected signal is around $\unit[50]{mHz}$) down-sampled to $\unit[1]{Hz}$ to ease data processing. During the mission, data will be collected at a sample rate between $1$ and $\unit[10]{Hz}$, depending on the experiment and available down-link bandwidth.

\subsection{Whitening filters} \label{sect:whitening_filters}

\begin{figure}[htb]
\includegraphics[width=\columnwidth]{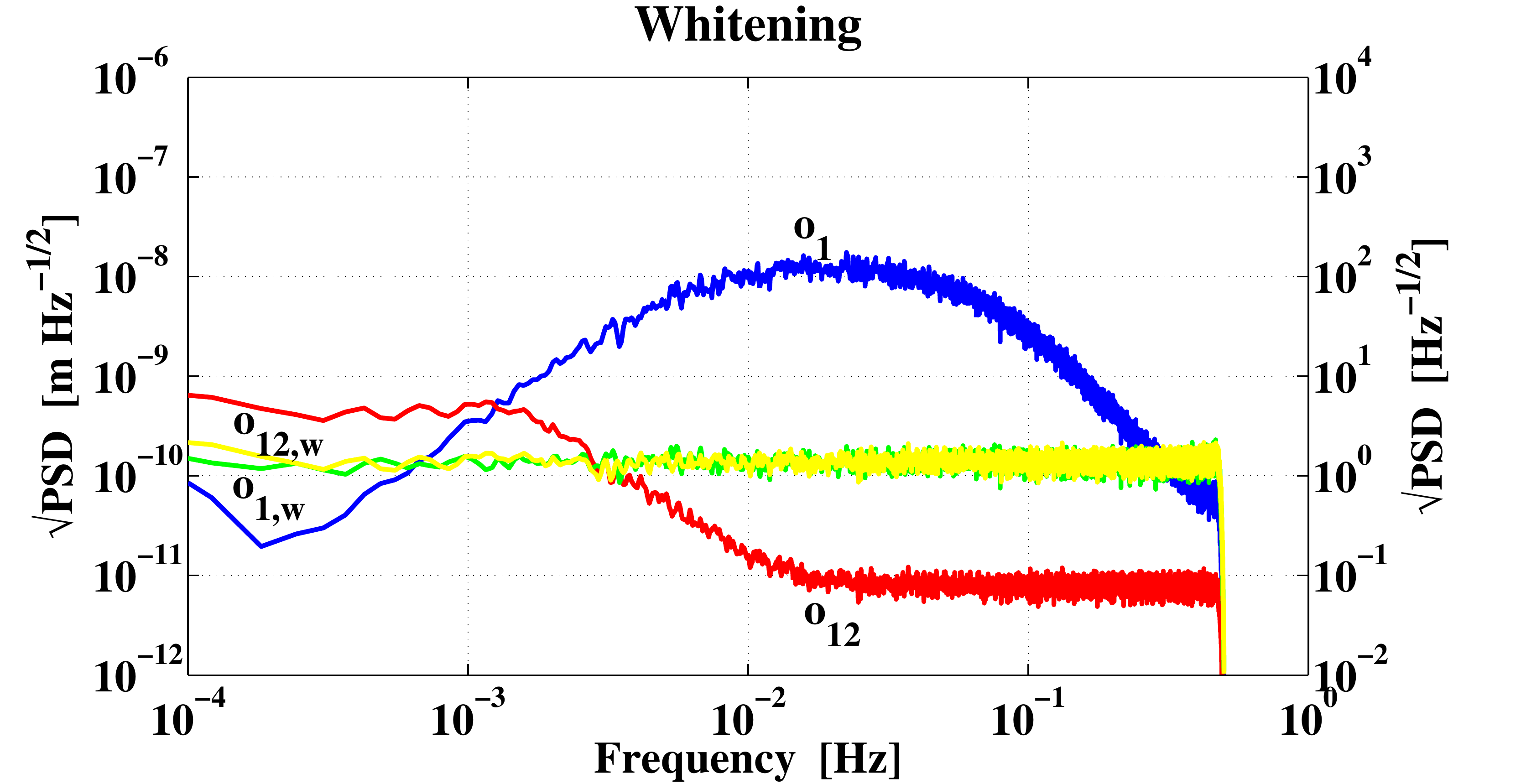}
\caption{\label{fig:whitening}\footnotesize{Effect of the whitening filters on a simulated noise run. $o_1$ and $o_{12}$ are the original noise stretches. $o_{1,\text{w}}$ and $o_{12,\text{w}}$ are the (flattened) whitened noise stretches. PSDs are computed with the Welch overlap method, 4-sample 92-dB Blackman-Harris window \cite{harris1978}, 16 averages and mean-detrending.}}
\end{figure}

The first step of the data analysis pipeline is the estimation of the whitening filters from the 28-hour noise run. For uncorrelated channels (in the parameter estimation experiments cross-correlation can be effectively ignored) whitening filters are derived following the principle of Section \ref{sect:whitening}: a fit in the $z$ domain to the inverse of the noise spectrum is performed. In Fig.\;\ref{fig:whitening} the resulting
effect of the whitening filters is to flatten the original noise curves. For each spectrum we report the estimated Power Spectral Density (PSD).

Other details are summarized in Table \ref{tab:whitening}. All statistical moments estimated on the whitened time-series are compatible with the ones expected according to Gaussian noise, except for the mean of the differential channel which shows a significant departure from the zero-mean value. This is due to an intrinsic limitation of the whitening process at low frequency.
\begin{table}[htb]
\squeezetable
\caption{\label{tab:whitening}\footnotesize{Sample mean $\mu$ and standard deviation $\sigma$ with higher moments, the sample skewness $\gamma_1$ and the excess kurtosis $\gamma_2$ for the whitened channels $o_1$ and $o_{12}$. Assuming Gaussian-distributed data, the approximate standard deviations are $\sigma_\mu\simeq\sigma/\sqrt{N}$, $\sigma_\sigma\simeq\sigma/\sqrt{2N}$, $\sigma_{\gamma_1}\simeq\sqrt{6/N}$, $\sigma_{\gamma_2}\simeq\sqrt{24/N}$, with $N$ the number of samples.}}
\begin{ruledtabular}
\begin{tabular}{ l D{!}{\,\pm\,}{5.5} D{!}{\,\pm\,}{5.5} D{!}{\,\pm\,}{2.6} D{!}{\,\pm\,}{2.6}}
                    & \multicolumn{1}{c}{$\mu$} & \multicolumn{1}{c}{$\sigma$}  & \multicolumn{1}{c}{$\gamma_1$} & \multicolumn{1}{c}{$\gamma_2$} \\
\hline
$o_1$               & 0.008 ! 0.003         & 0.970 ! 0.002     & (\minus5 ! 8)$\e{\minus3}$    & (0 ! 2)$\e{\minus2}$ \\
$o_{12}$            & \minus0.254 ! 0.003   & 1.002 ! 0.002     & (0 ! 8)$\e{\minus3}$          & (3 ! 2)$\e{\minus2}$ \\
\end{tabular}
\end{ruledtabular}
\end{table}

\subsection{Parameter estimation} \label{sect:param_est}

\begin{figure}[htb]
\includegraphics[width=\columnwidth]{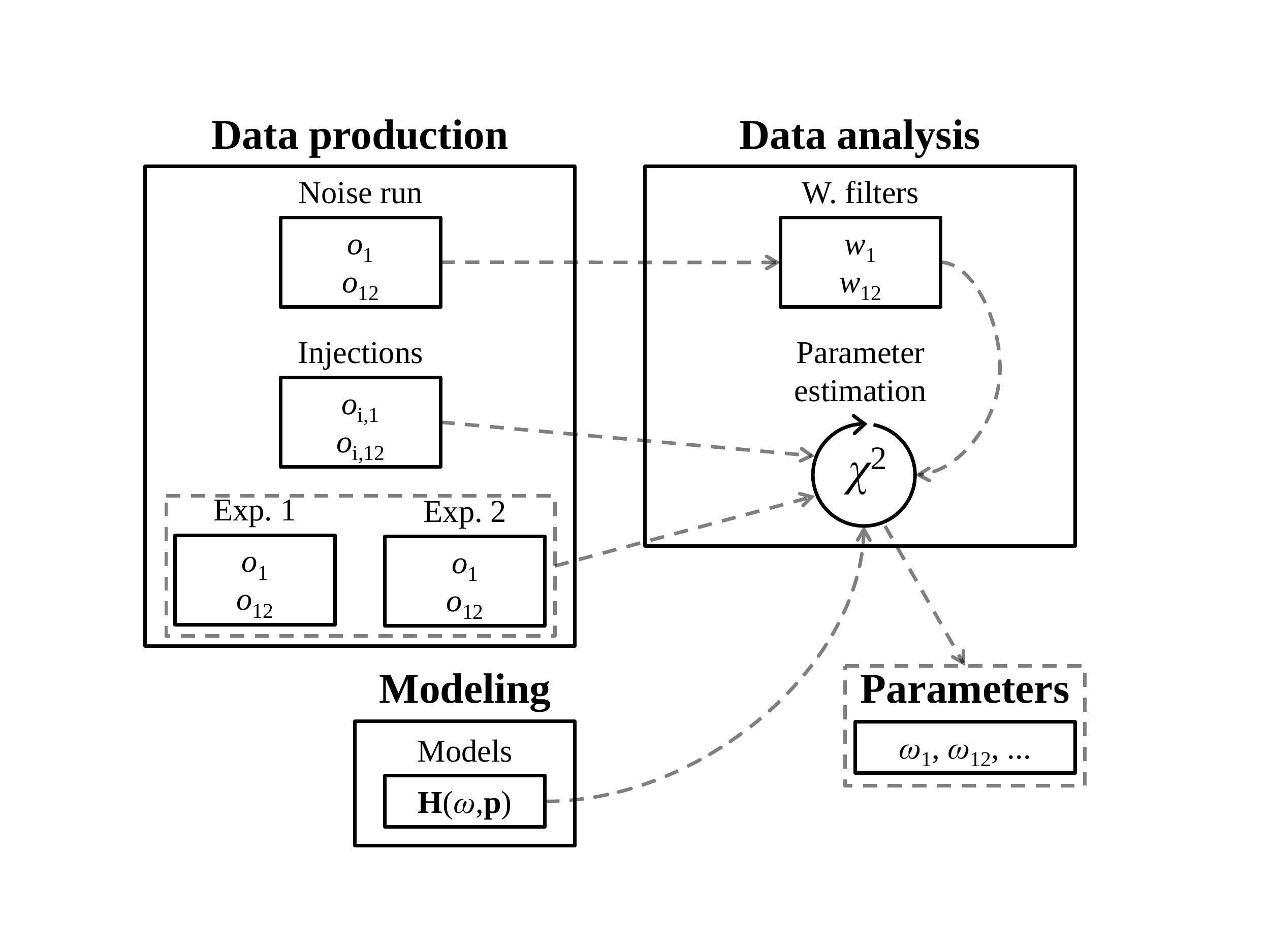}
\vspace{-30pt}
\caption{\label{fig:param_est_scheme}\footnotesize{A simplified schematic of the parameter estimation algorithm. After data production, the long noise run is used to estimate the whitening filters. Then, all injections and system responses are collected and passed to the $\chi^2$ fitting algorithm. The modeling provides the matrix of transfer functions. By adjusting the parameter the $\chi^2$ is optimized to get the best-fit parameters.}}
\end{figure}

In what follows we implement the ideas of Section \ref{sect:maximum_likelihood} on
simulated data. The objective function is the log-likelihood ($\chi^2$ statistic
in least-squares sense) which is optimized with respect to the parameter
vector. The method uses all channels and experiments to maximize the information
and remove the degeneracy. Moreover, all models are treated non-linearly. The matrix of the transfer
function models, the inputs and outputs for each experiment, together with
the whitening filters for all channels are passed to the fitting algorithm. See Fig.\;\ref{fig:param_est_scheme} for an idealized and simplified scheme that shows how the data and models are used by the parameter estimation algorithm.

The optimization approach is mixed: a preliminary search with the preconditioned conjugate gradient algorithm (alternatively, the Broyden-Fletcher-Goldfarb-Shanno quasi-Newton method can be used) \cite{press2007}, using analytical derivatives, is followed by a derivative-free
simplex algorithm. The main advantage is that a global minimizer with lower
precision is accompanied by a local minimizer with higher precision and this overcomes
the intrinsic difficulties connected with non-linear optimizations.

\subsubsection{Monte Carlo validation} \label{sect:monte_carlo}

A Monte Carlo simulation with $1000$ different noise realizations was used to check for consistency of the method. The estimation was identically repeated at each step, enabling fine tuning and the study of the statistics for every relevant physical quantity. In Table \ref{tab:montecarlo} we compare the mean best-fit to the true values for all parameters: the accordance is at the level of $1$ or $2$ standard deviations. We also show the sample standard deviation of the best-fits and the mean estimated standard deviation: the first one is the fluctuation of the parameters due to the noise, the second is the average fit error.
\begin{table}[htb]
\squeezetable
\caption{\label{tab:montecarlo}\footnotesize{Monte Carlo validation of $1000$ independent noise realizations. The mean best-fits are compatible with the real values. The term in brackets is the error relative to the rightmost digit. The mean standard deviation (estimated from the fit) is of the same order of magnitude of the sample standard deviation of the best-fits. The mean $\chi^2=0.96$, $\nu=79993$.}}
\begin{ruledtabular}
\begin{tabular}{l D{.}{.}{1.3} D{.}{.}{1.7} c c}
& \multicolumn{1}{l}{\multirow{2}{*}{Real}} & \multicolumn{1}{c}{Mean} & St.\;dev.\;of & Mean \\
& & \multicolumn{1}{c}{best-fit} & \multicolumn{1}{c}{best-fit} &  st.\;dev. \\
\hline
$A_\mathrm{df}$                                     & 1.003          & 1.00297(1)           & 4\e{\minus4}  & 4\e{\minus4} \\
$A_\mathrm{sus}$                                    & 0.9999         & 0.9999001(1)         & 4\e{\minus6}  & 2\e{\minus5} \\
$S_{21}\,[10^{\minus4}]$                            & 0.9            & 0.90004(9)           & 3\e{\minus3}  & 4\e{\minus3} \\
$\omega_1^2\,[10^{\minus6}\,\text{s}^{\minus2}]$    & \minus1.303    & \minus1.303006(7)    & 2\e{\minus4}  & 1\e{\minus3} \\
$\omega_{12}^2\,[10^{\minus6}\,\text{s}^{\minus2}]$ & \minus0.698    & \minus0.697998(6)    & 2\e{\minus4}  & 5\e{\minus4} \\
$\Delta t_1\,[\text{s}]$                            & 0.06           & 0.059995(3)          & 9\e{\minus5}  & 3\e{\minus4} \\
$\Delta t_{12}\,[\text{s}]$                         & 0.05           & 0.05000(3)           & 8\e{\minus4}  & 1\e{\minus3} \\
\end{tabular}
\end{ruledtabular}
\end{table}

\begin{figure*}[htb]
\begin{tabular}{cccc}
\hspace{-10pt} \resizebox{47mm}{!}{\includegraphics{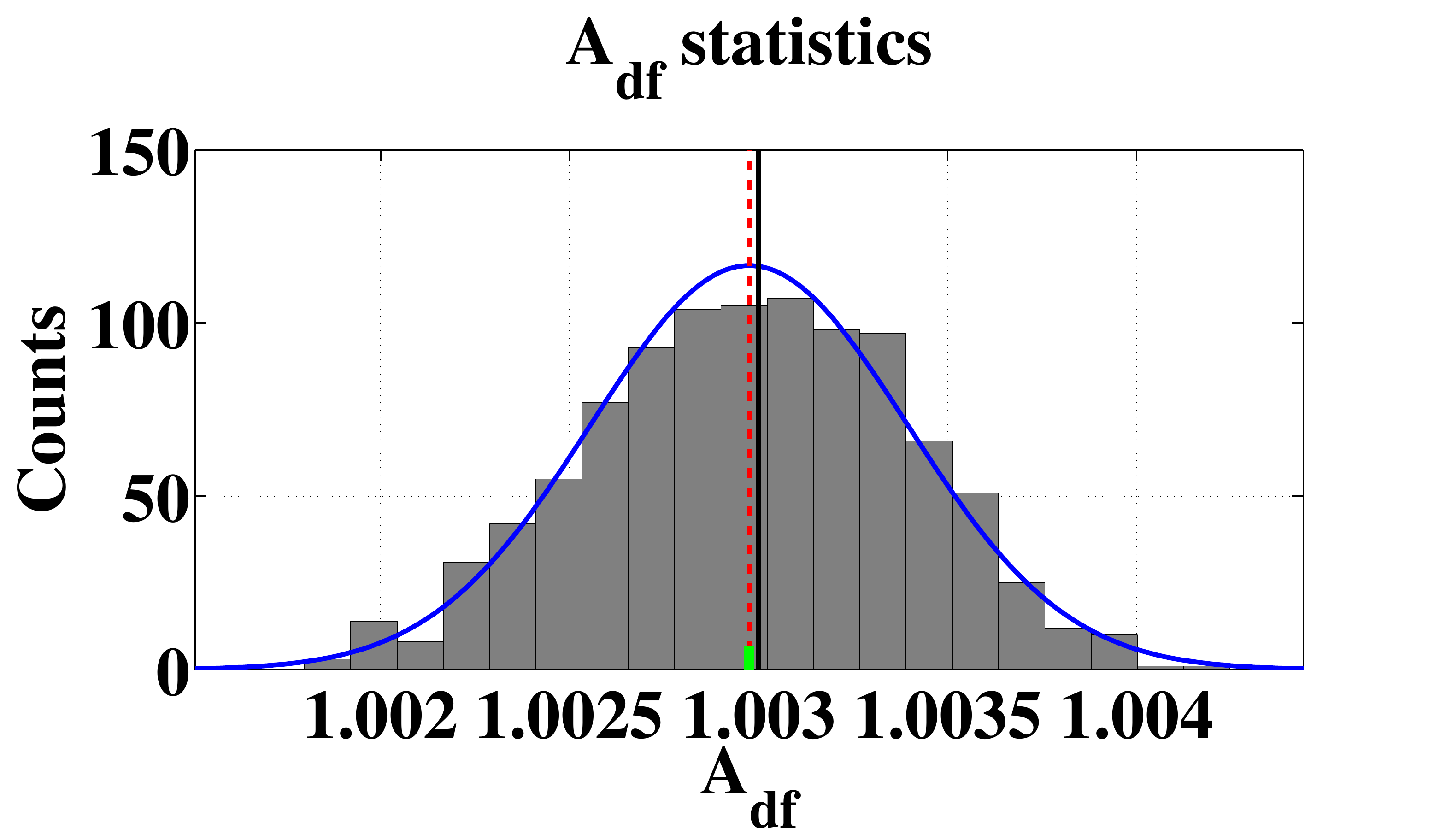}} &
\hspace{-15pt} \resizebox{47mm}{!}{\includegraphics{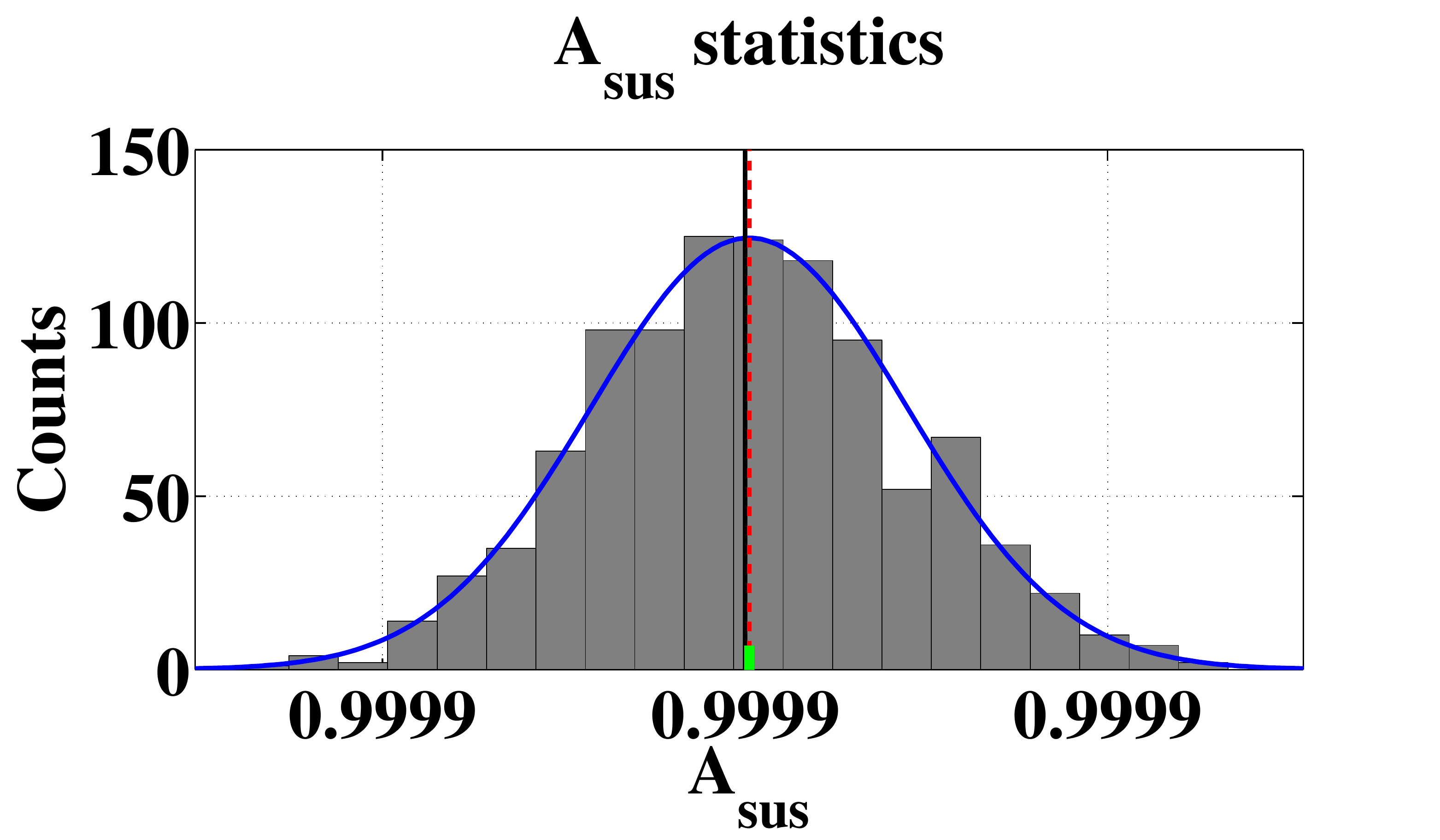}} &
\hspace{-15pt} \resizebox{47mm}{!}{\includegraphics{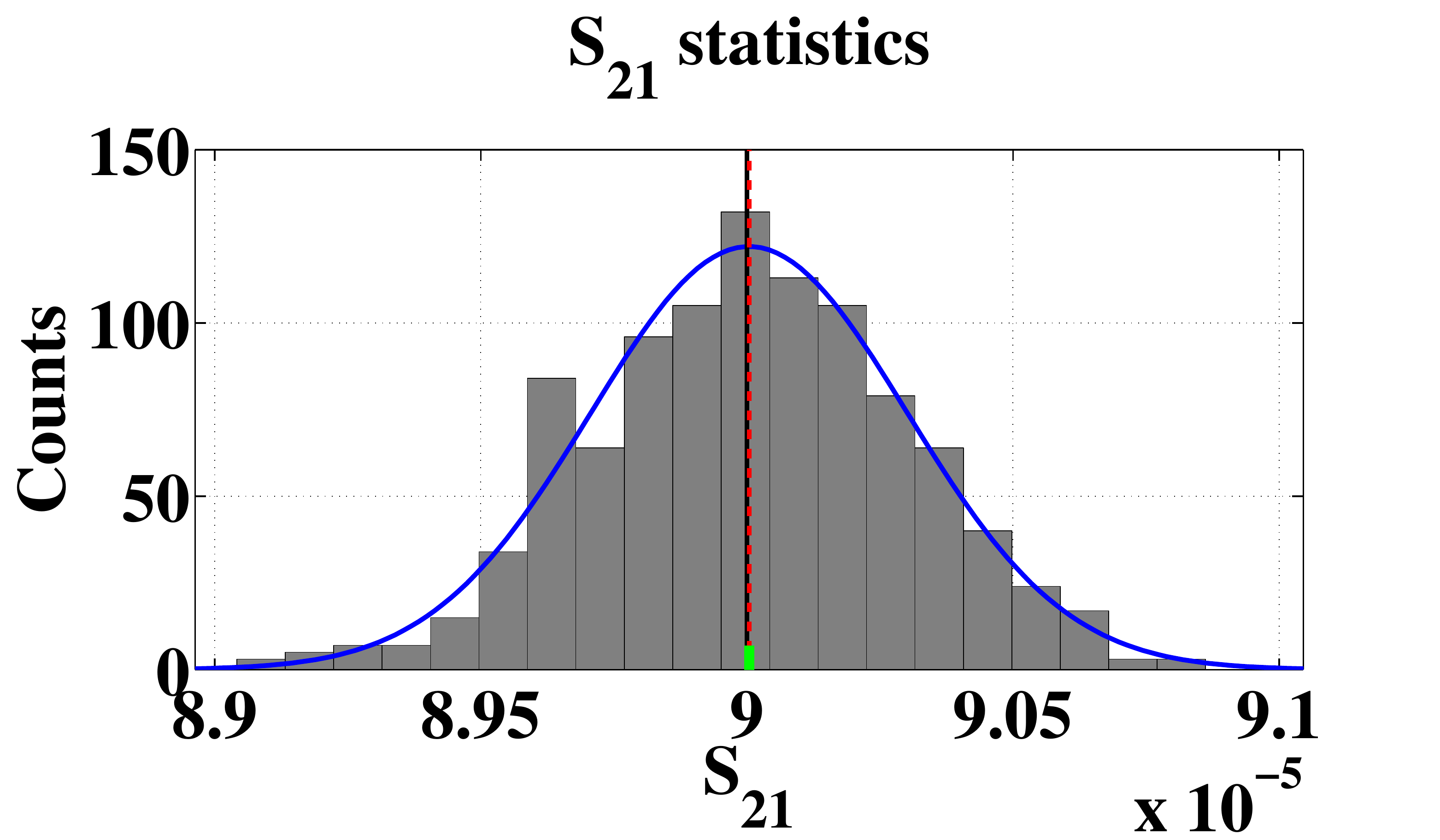}} &
\hspace{-15pt} \resizebox{47mm}{!}{\includegraphics{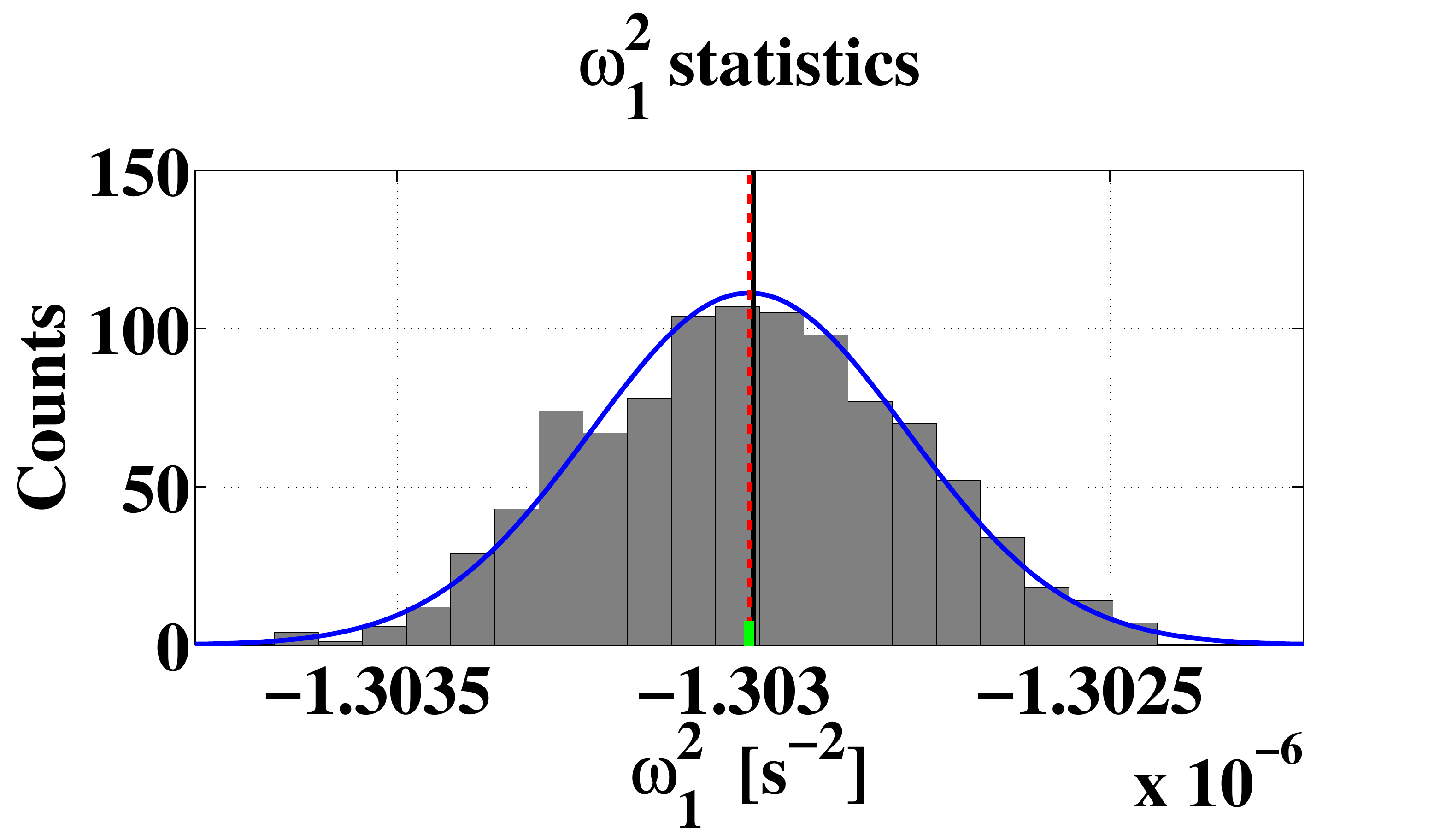}} \\
\hspace{-10pt} (a) & \hspace{-15pt} (b) & \hspace{-15pt} (c) & \hspace{-15pt} (d) \\
\vspace{-5pt}\\
\hspace{-10pt} \resizebox{47mm}{!}{\includegraphics{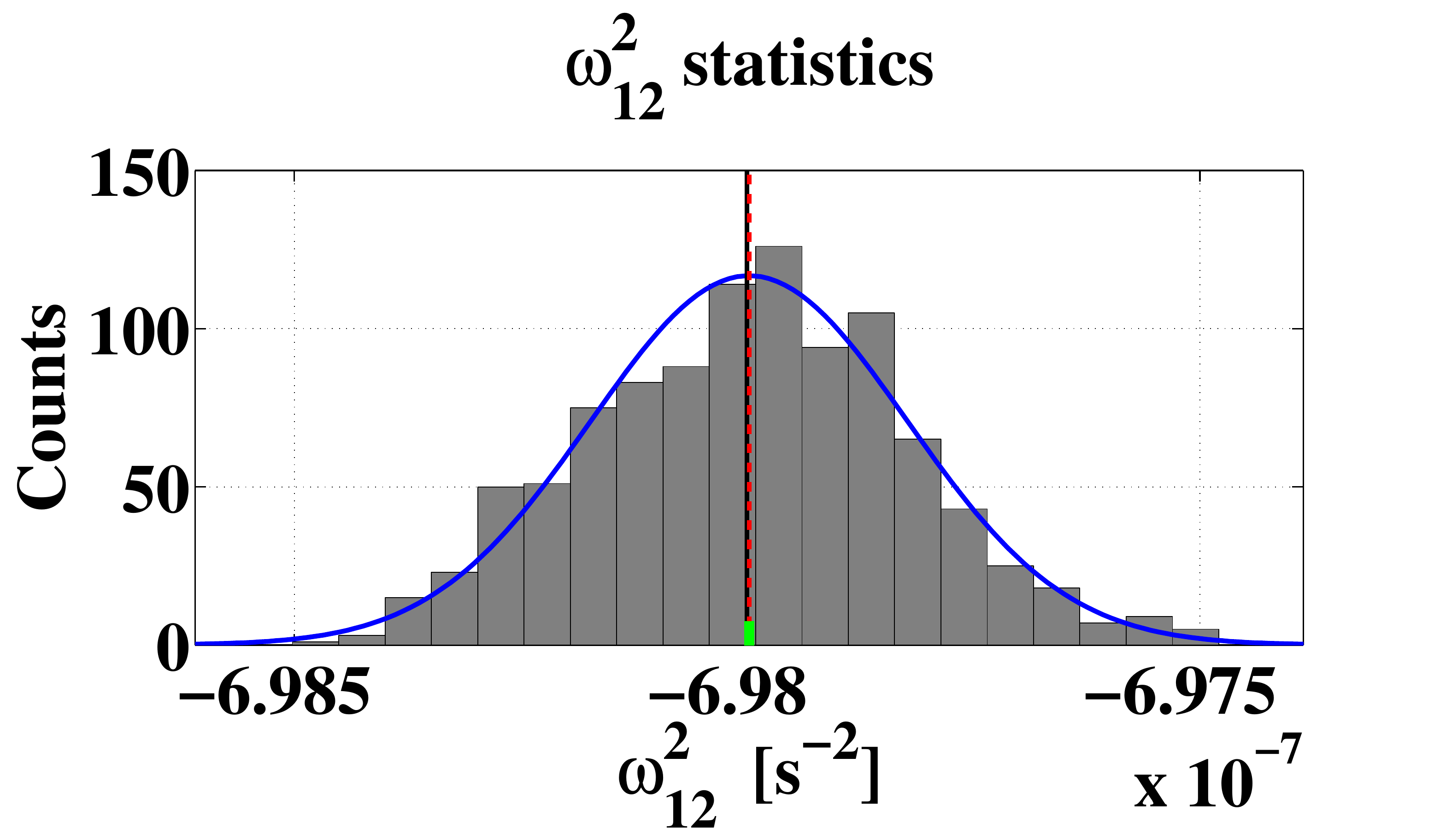}} &
\hspace{-15pt} \resizebox{47mm}{!}{\includegraphics{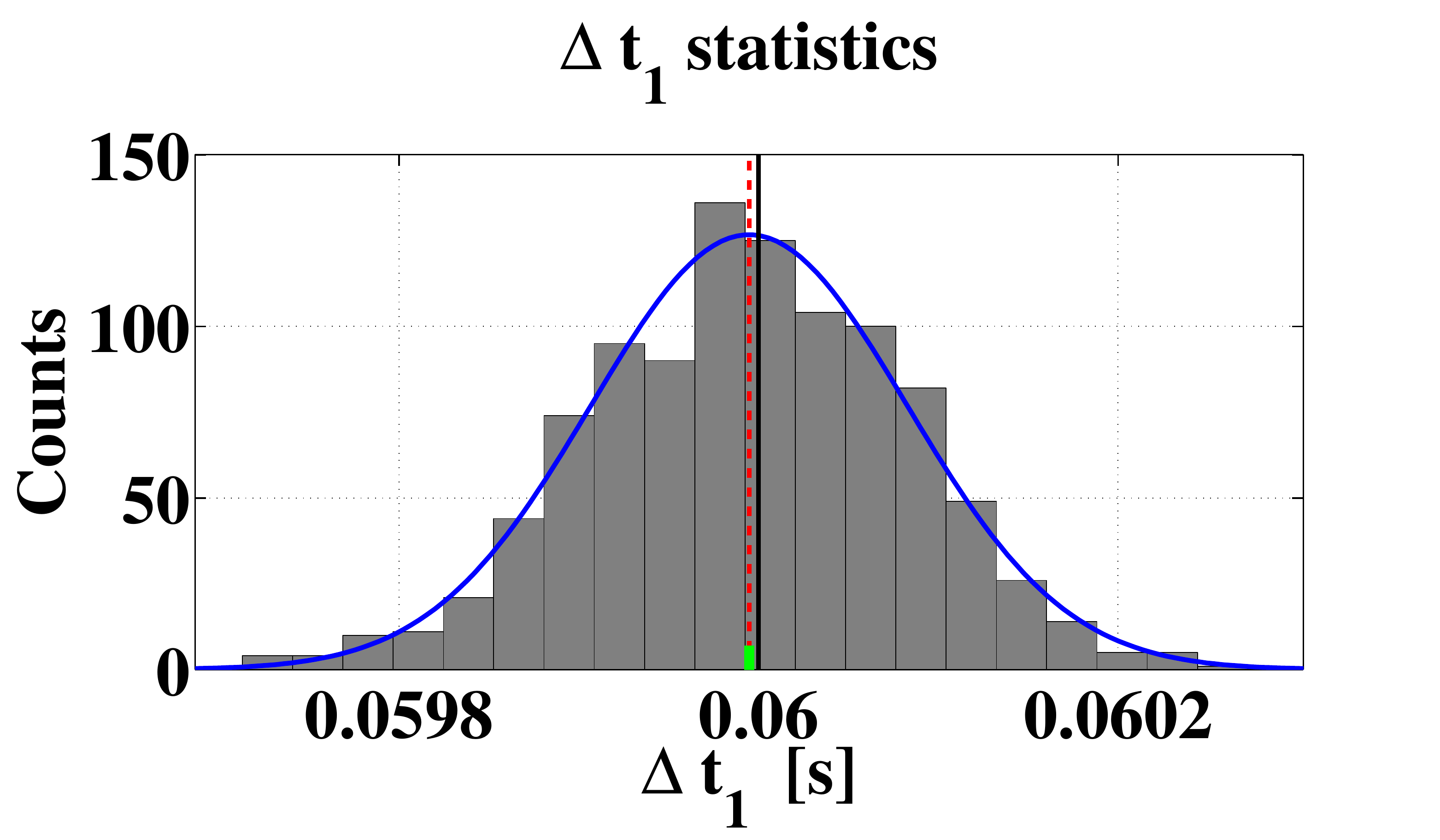}} &
\hspace{-15pt} \resizebox{47mm}{!}{\includegraphics{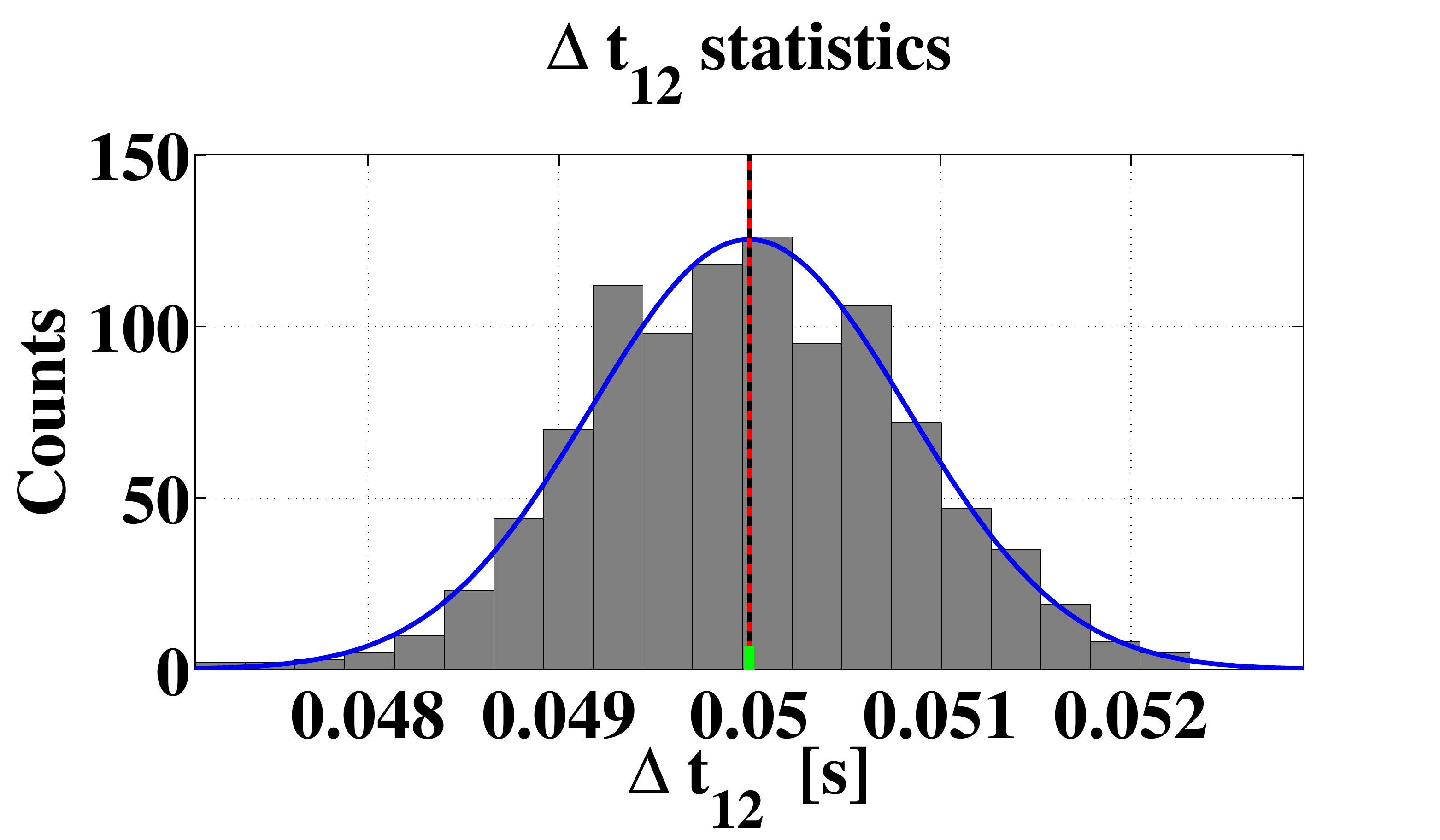}} \\
\hspace{-10pt} (e) & \hspace{-15pt} (f) & \hspace{-15pt} (g) \\
\end{tabular}
\caption{\label{fig:montecarlo_parameters}\footnotesize{Monte Carlo statistics for each parameter (a)-(g). The scaled Gaussian PDF is evaluated at the sample mean (dashed vertical line) and sample standard deviation (half horizontal bar), to be compared to the true value (solid vertical line).}}
\end{figure*}

A more in-depth analysis concerns on the parameter statistics reported in Fig.\;\ref{fig:montecarlo_parameters}. The accordance of the sample statistics with the theoretical Gaussian Probability Density Function (PDF) (evaluated at the sample mean and standard deviation) is self-evident.

\begin{figure*}[htb]
\begin{tabular}{cccc}
\hspace{-10pt} \resizebox{47mm}{!}{\includegraphics{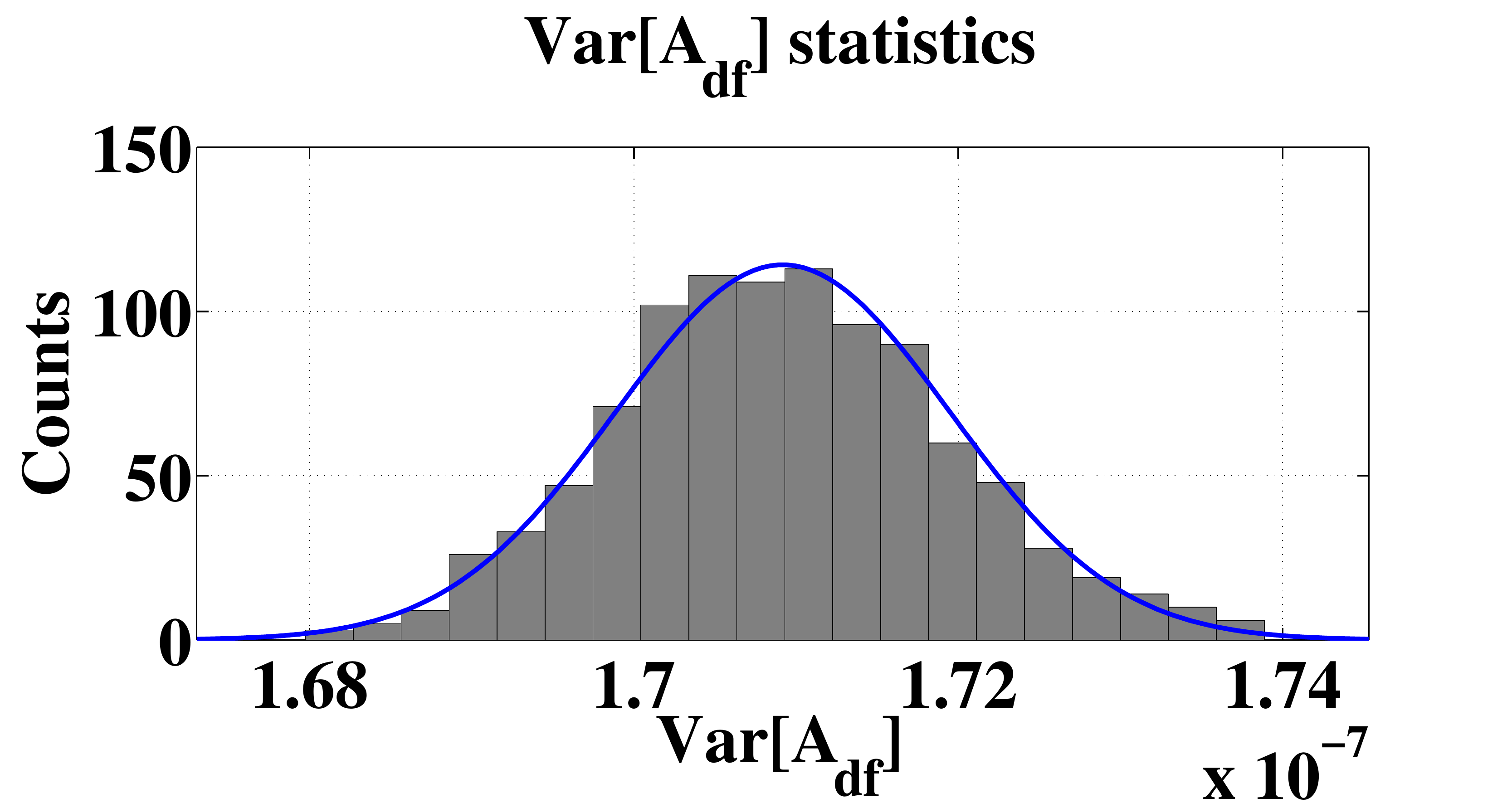}} &
\hspace{-15pt} \resizebox{47mm}{!}{\includegraphics{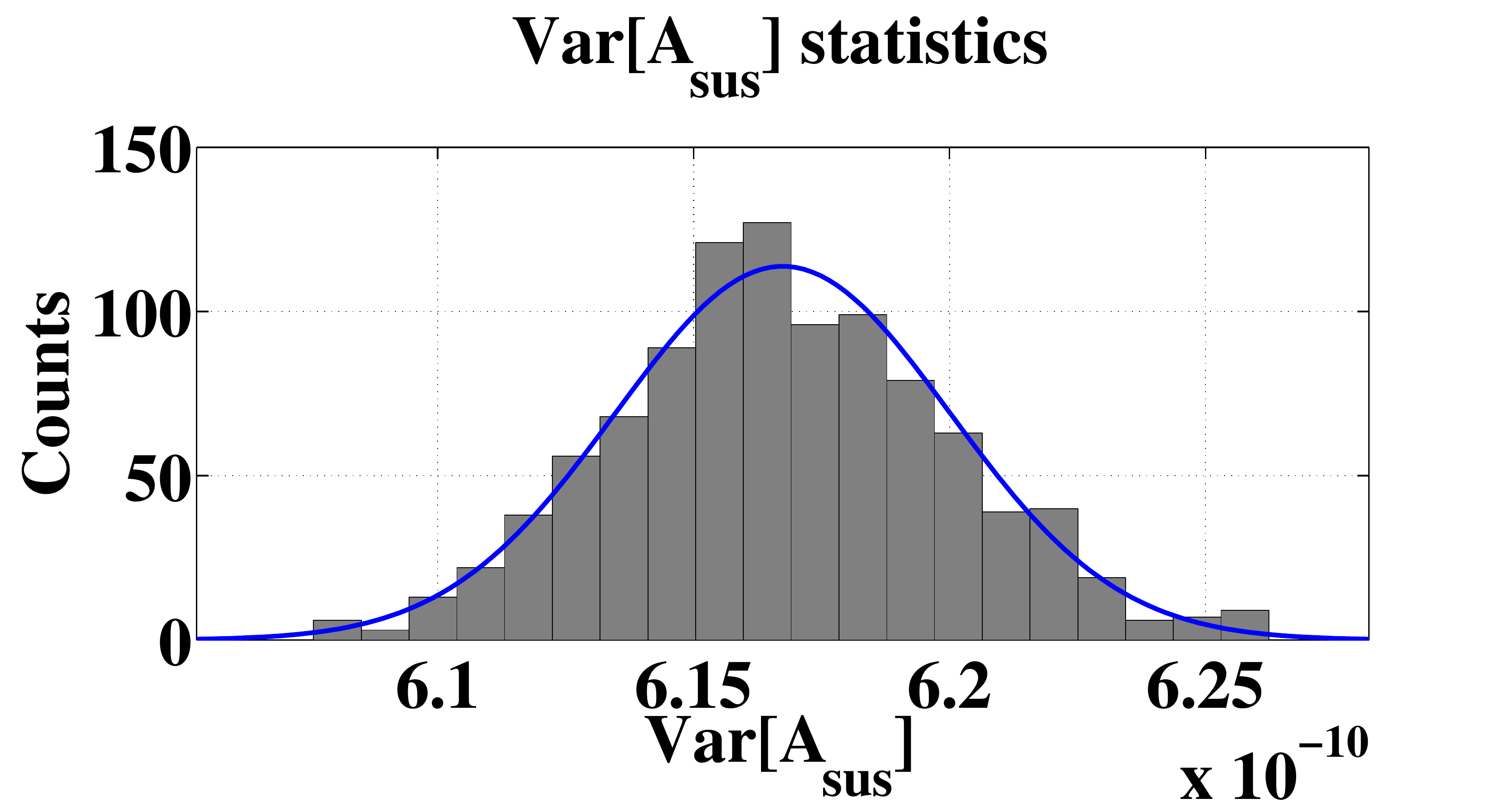}} &
\hspace{-15pt} \resizebox{47mm}{!}{\includegraphics{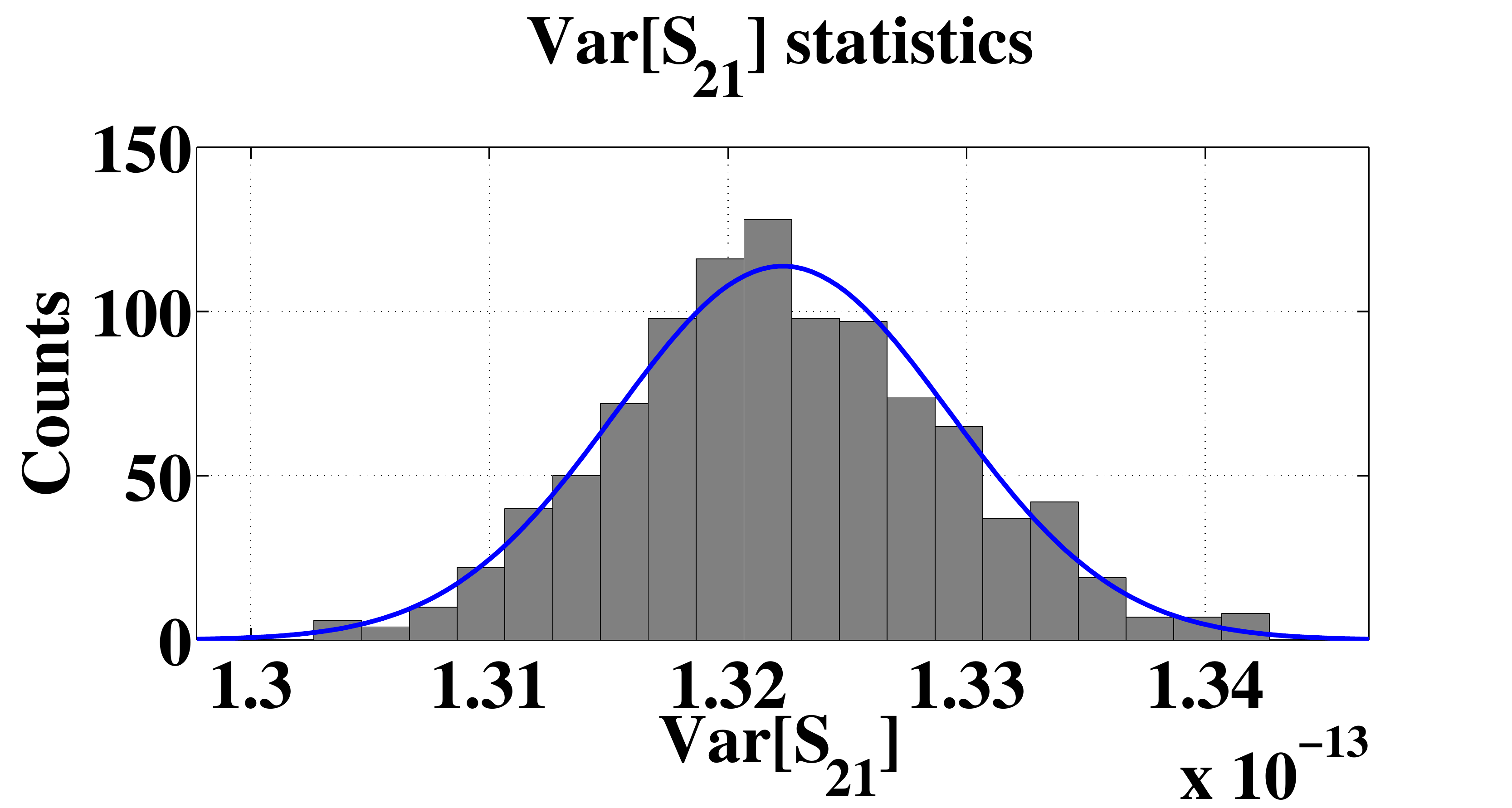}} &
\hspace{-15pt} \resizebox{47mm}{!}{\includegraphics{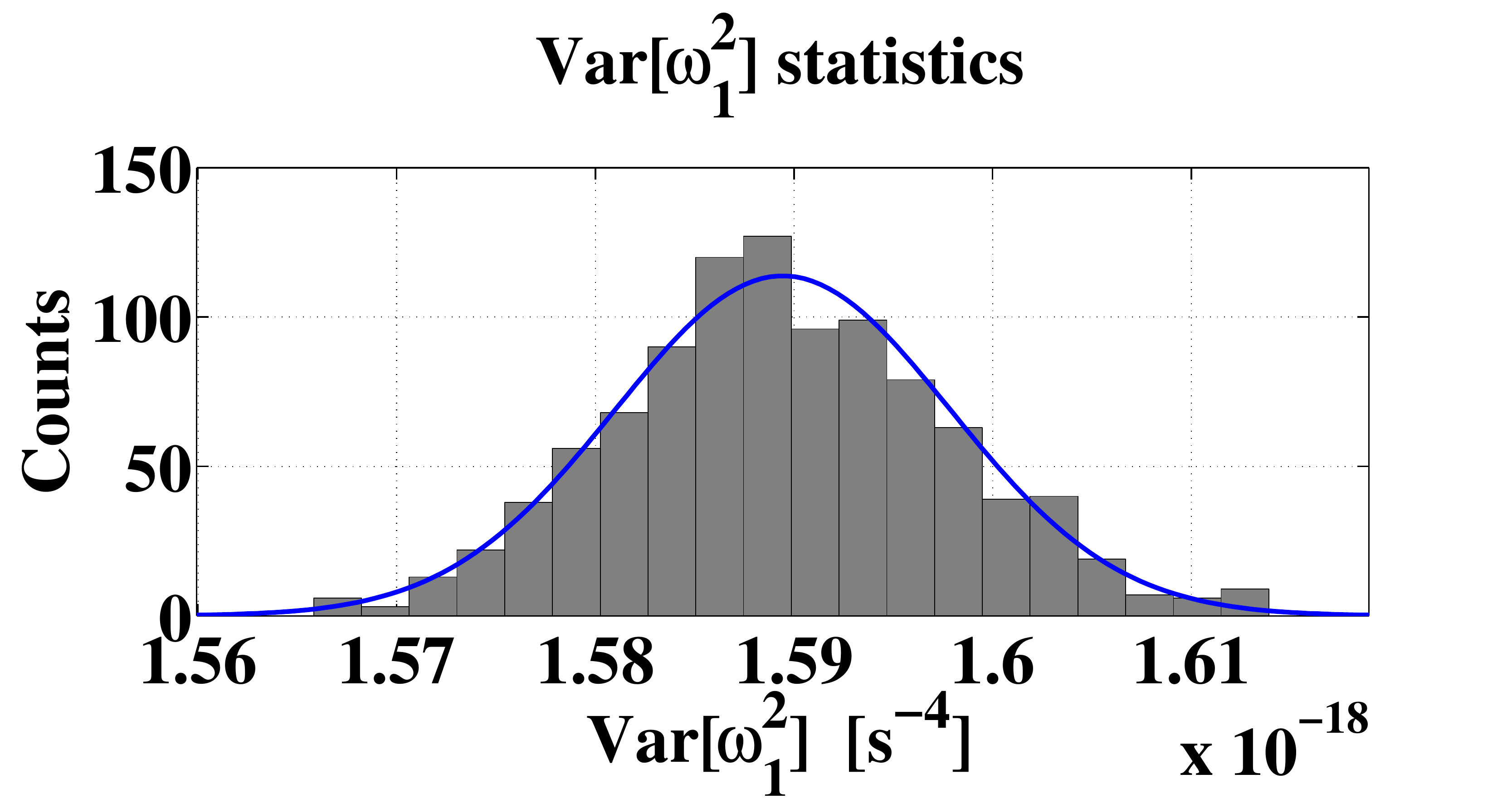}} \\
\hspace{-10pt} (a) & \hspace{-15pt} (b) & \hspace{-15pt} (c) & \hspace{-15pt} (d) \\
\vspace{-5pt}\\
\hspace{-10pt} \resizebox{47mm}{!}{\includegraphics{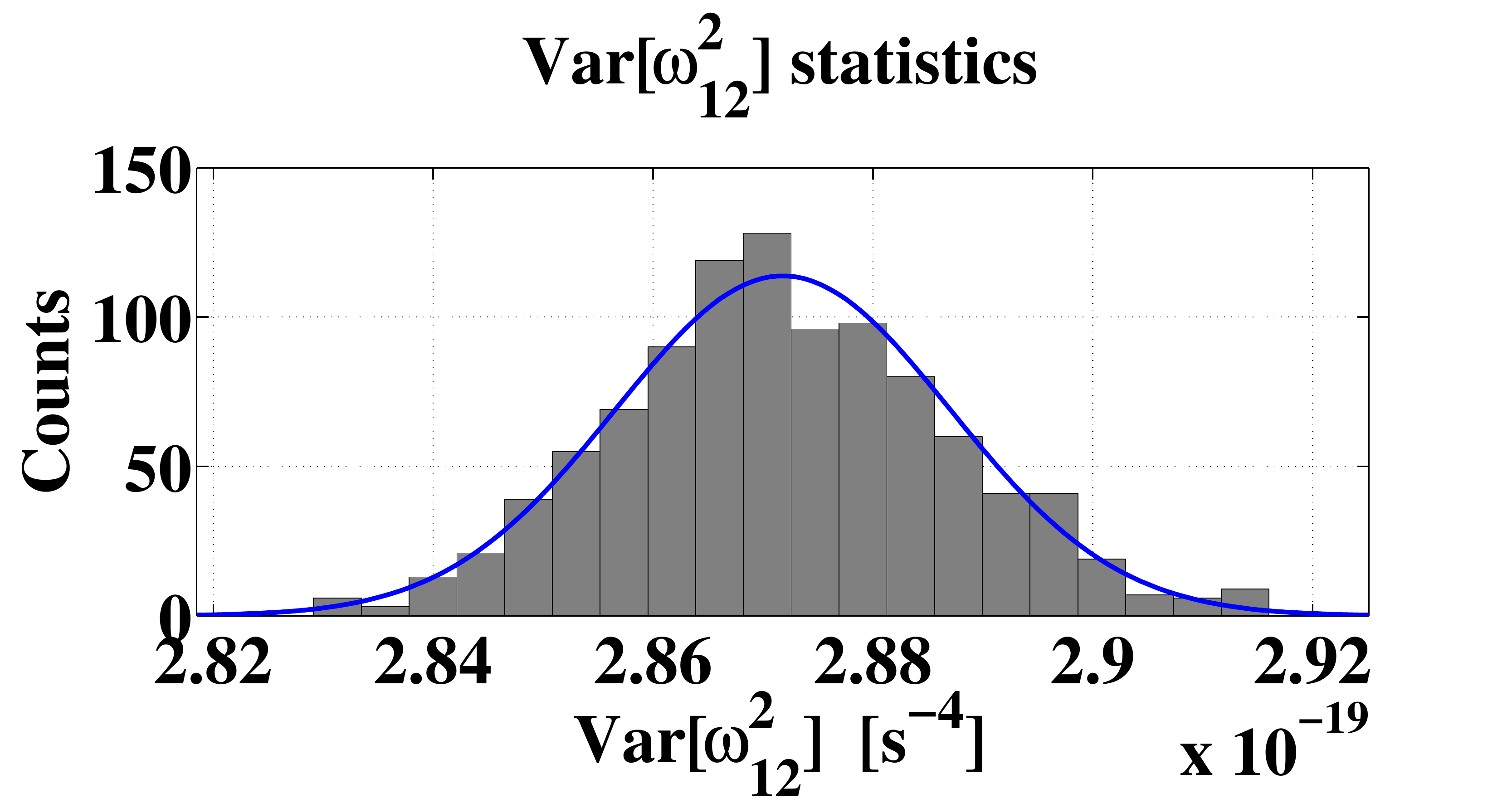}} &
\hspace{-15pt} \resizebox{47mm}{!}{\includegraphics{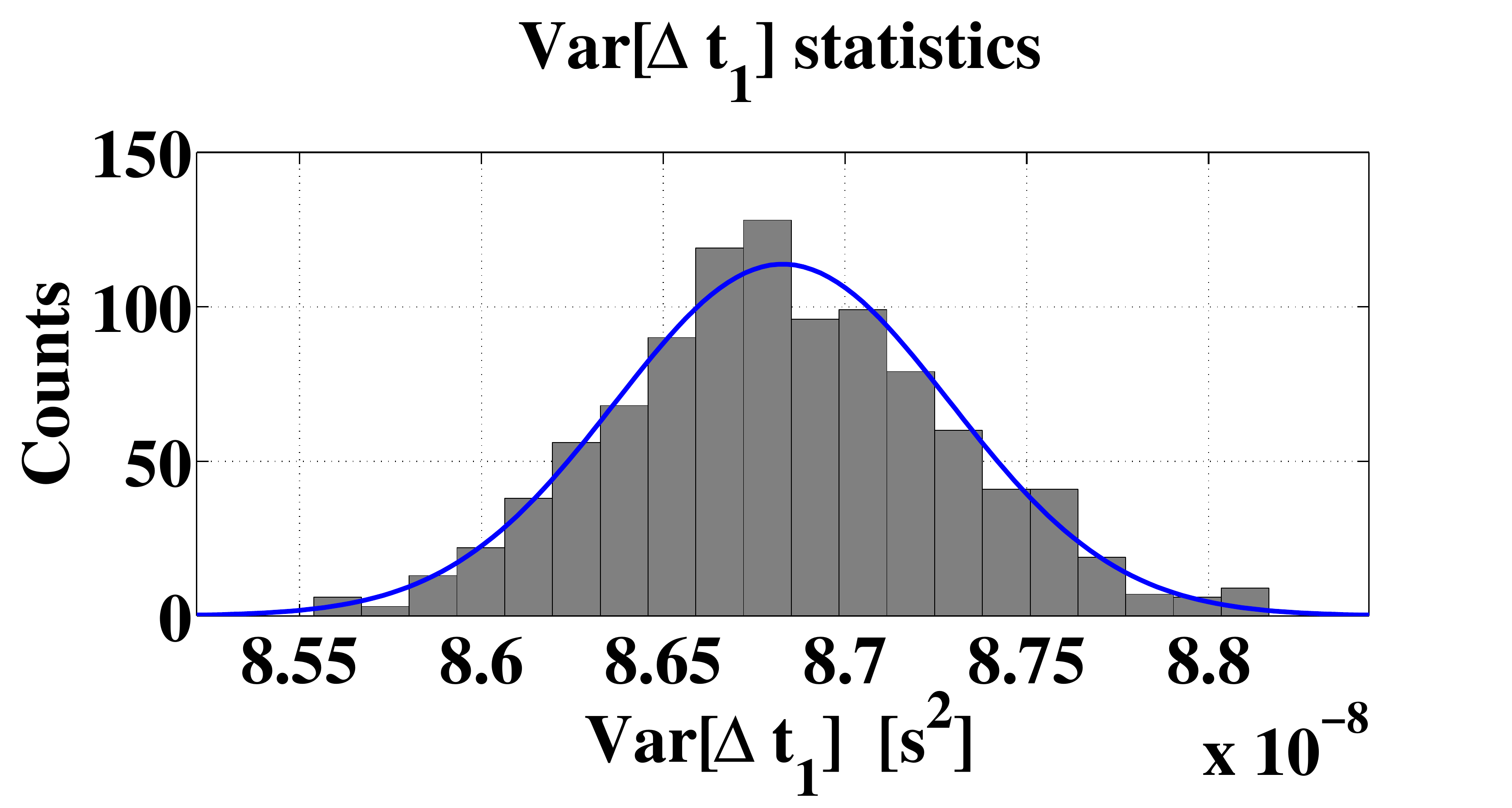}} &
\hspace{-15pt} \resizebox{47mm}{!}{\includegraphics{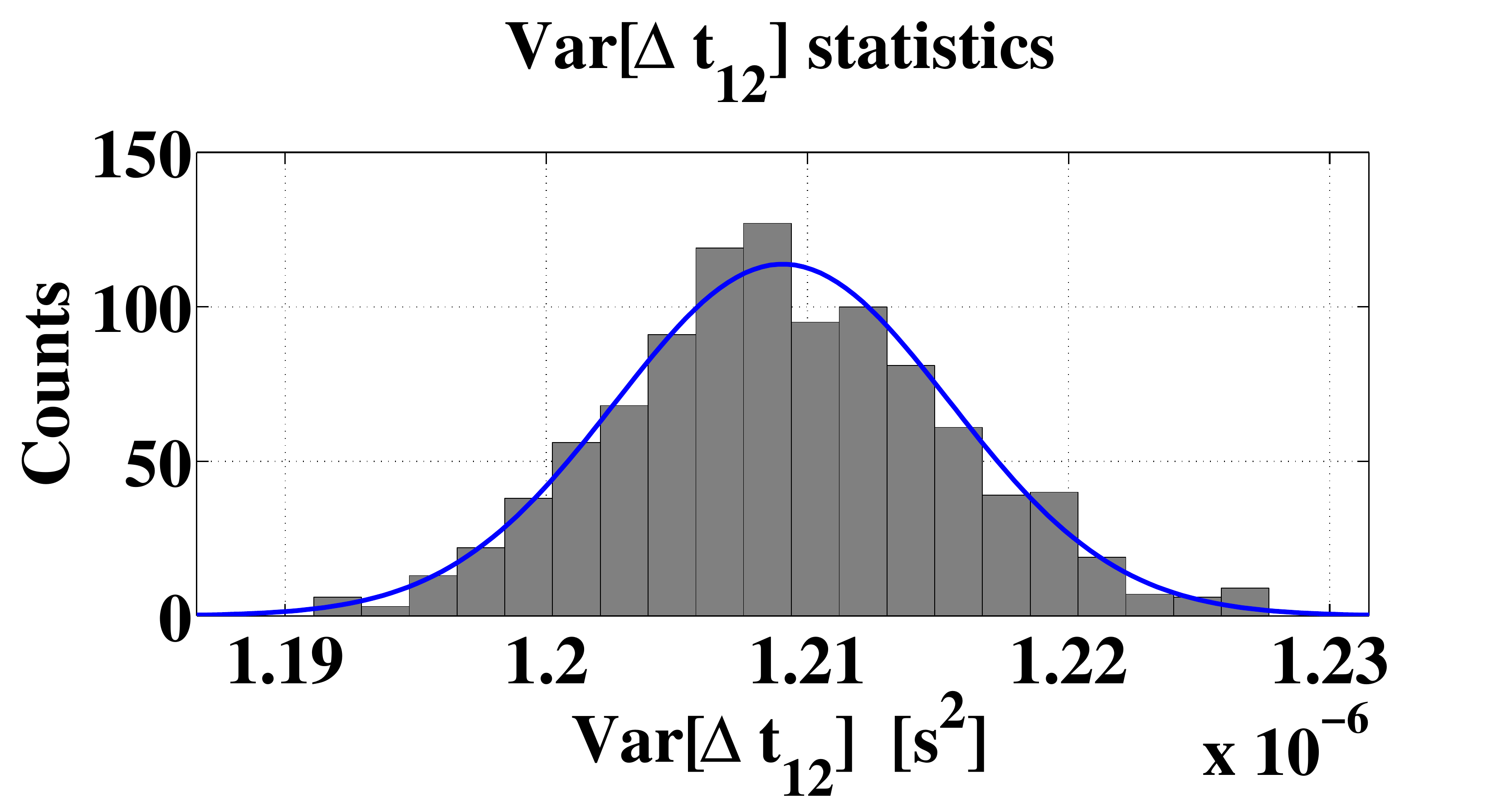}} \\
\hspace{-10pt} (e) & \hspace{-15pt} (f) & \hspace{-15pt} (g) \\
\end{tabular}
\caption{\label{fig:montecarlo_uncertainties}\footnotesize{Monte Carlo statistics for each parameter variance (a)-(g). The scaled Gaussian PDF is evaluated at the sample mean and standard deviation.}}
\end{figure*}

As for all parameters, in Fig.\;\ref{fig:montecarlo_uncertainties} we show the statistics for the estimated variances. Theory prescribes that the variance statistics should be $\chi^2$ distributed, but for $\nu=79993$ the $\chi^2$ distribution tends to a Gaussian distribution. From the plot it is clear that they are all in agreement with the theoretical Gaussian PDF.

\begin{figure}[htb]
\begin{tabular}{cc}
\includegraphics[width=\columnwidth]{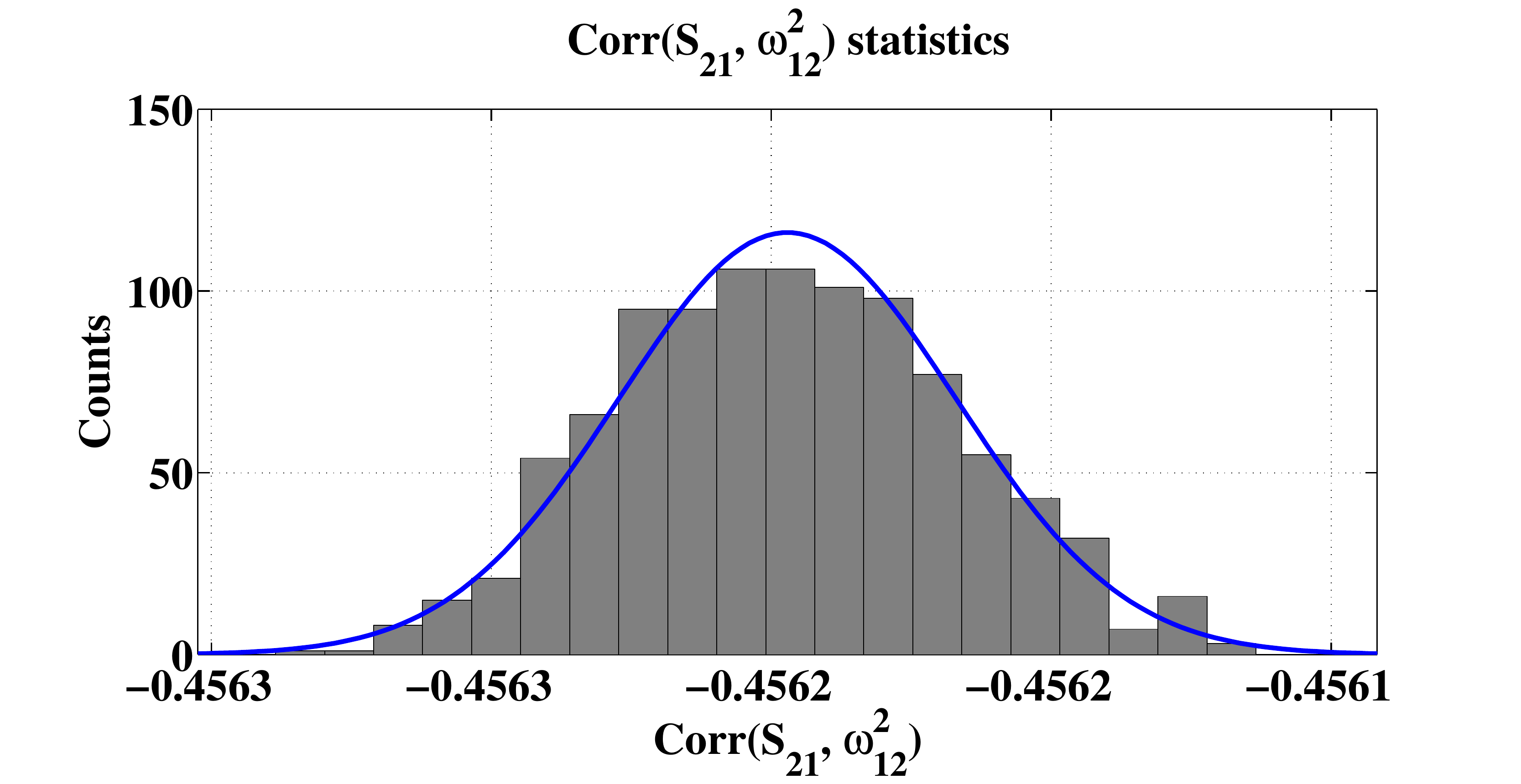} \\
\includegraphics[width=\columnwidth]{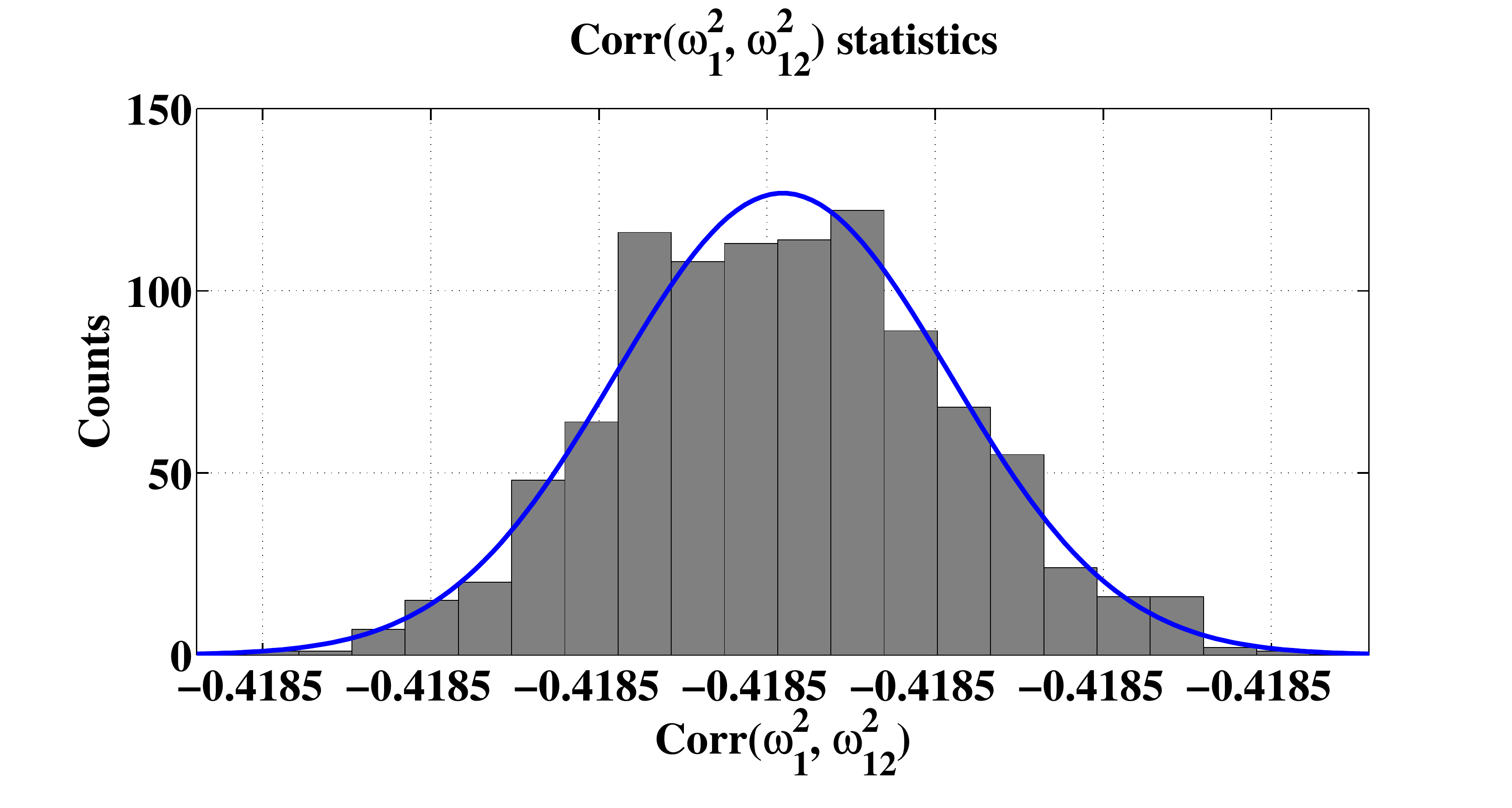}
\end{tabular}
\caption{\label{fig:montecarlo_corr}\footnotesize{Monte Carlo statistics for two parameter correlations. The scaled Gaussian PDF is evaluated at the sample mean and standard deviation.}}
\end{figure}

Surprisingly, the correlations are also Gaussian distributed with good approximation. See Fig.\;\ref{fig:montecarlo_corr} for two examples.

\begin{figure*}[htb]
\begin{tabular}{cc}
\resizebox{73mm}{!}{\includegraphics{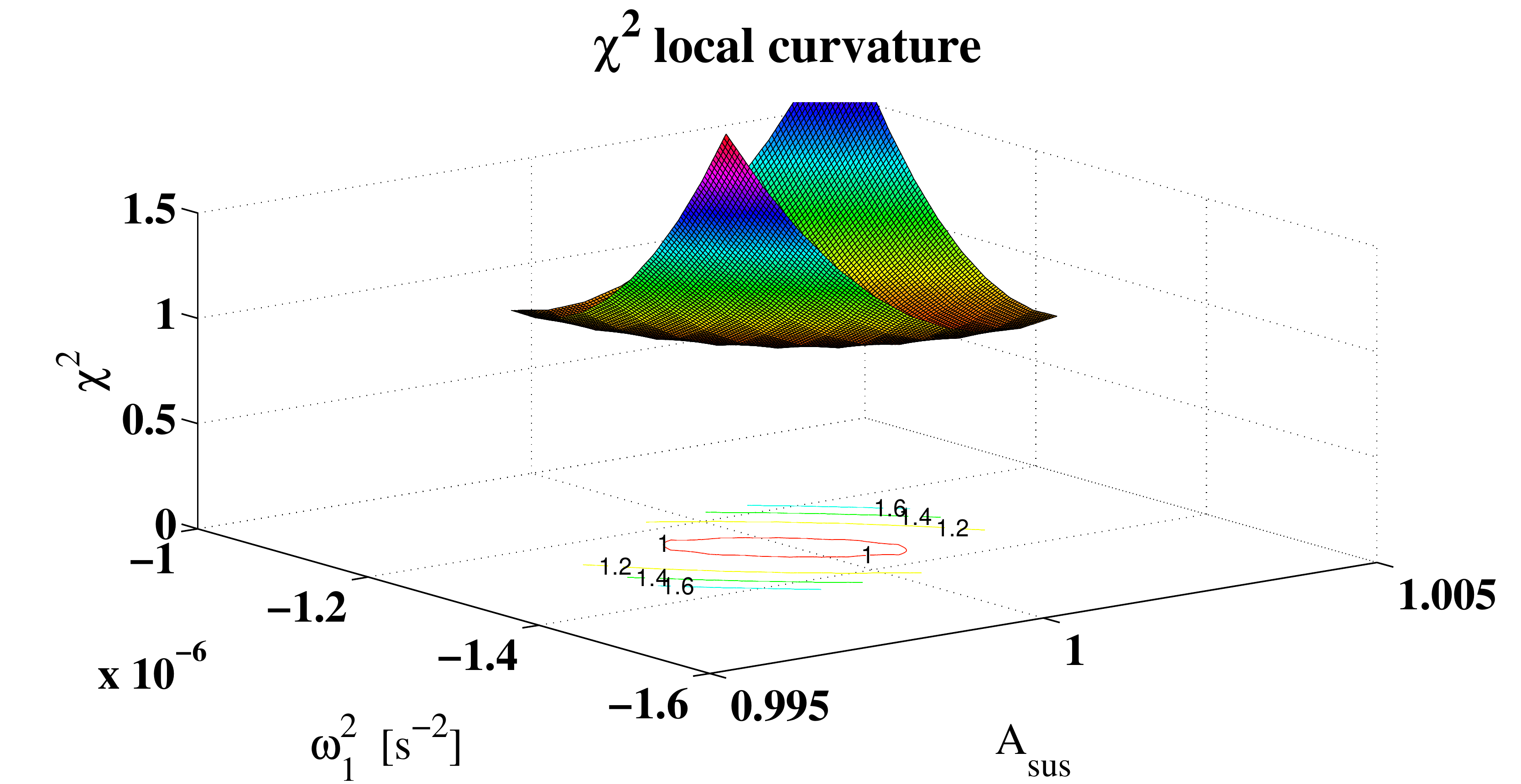}} &
\resizebox{73mm}{!}{\includegraphics{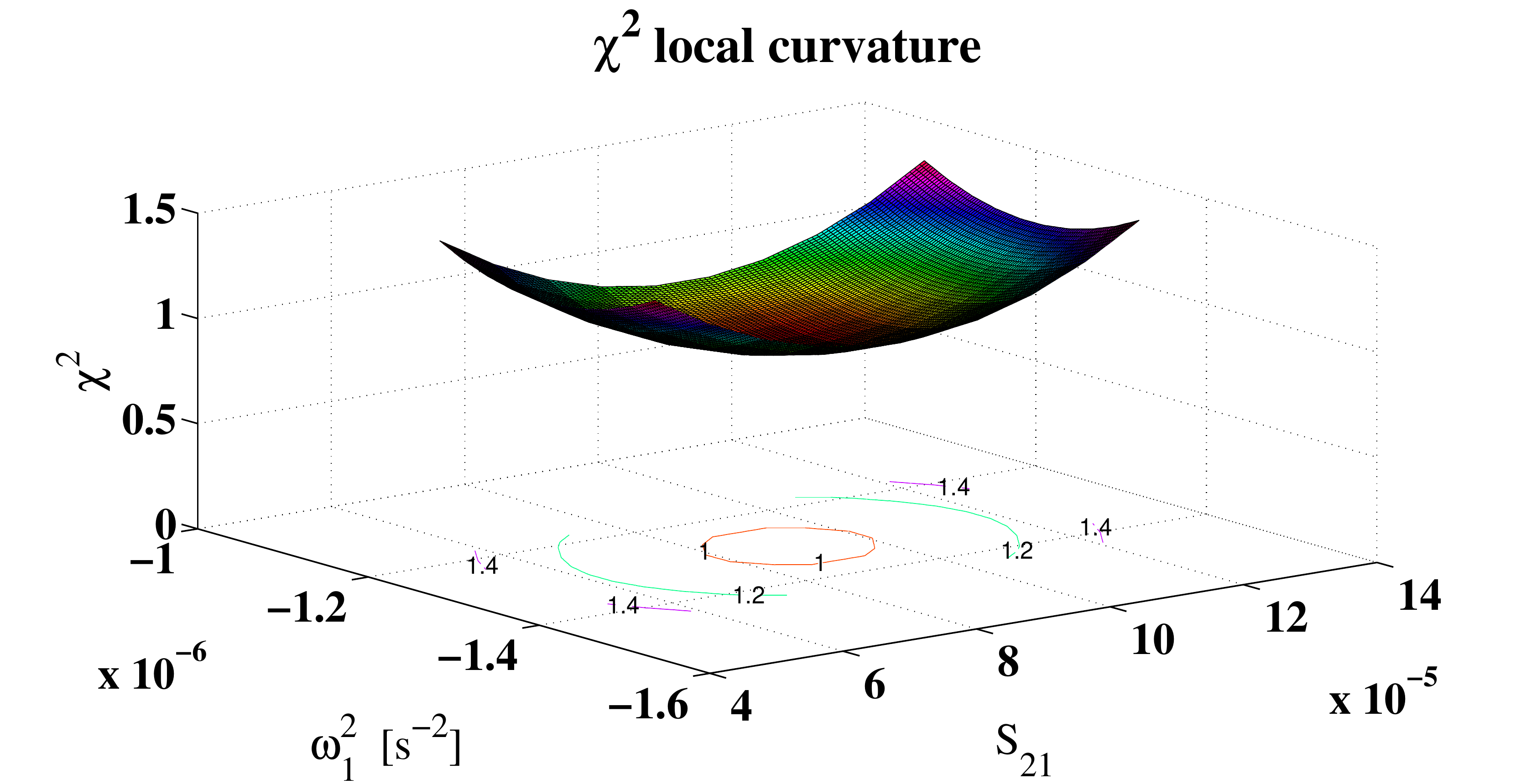}} \\
\hspace{5pt} (a) & \hspace{5pt} (b) \\
\vspace{-5pt} \\
\resizebox{73mm}{!}{\includegraphics{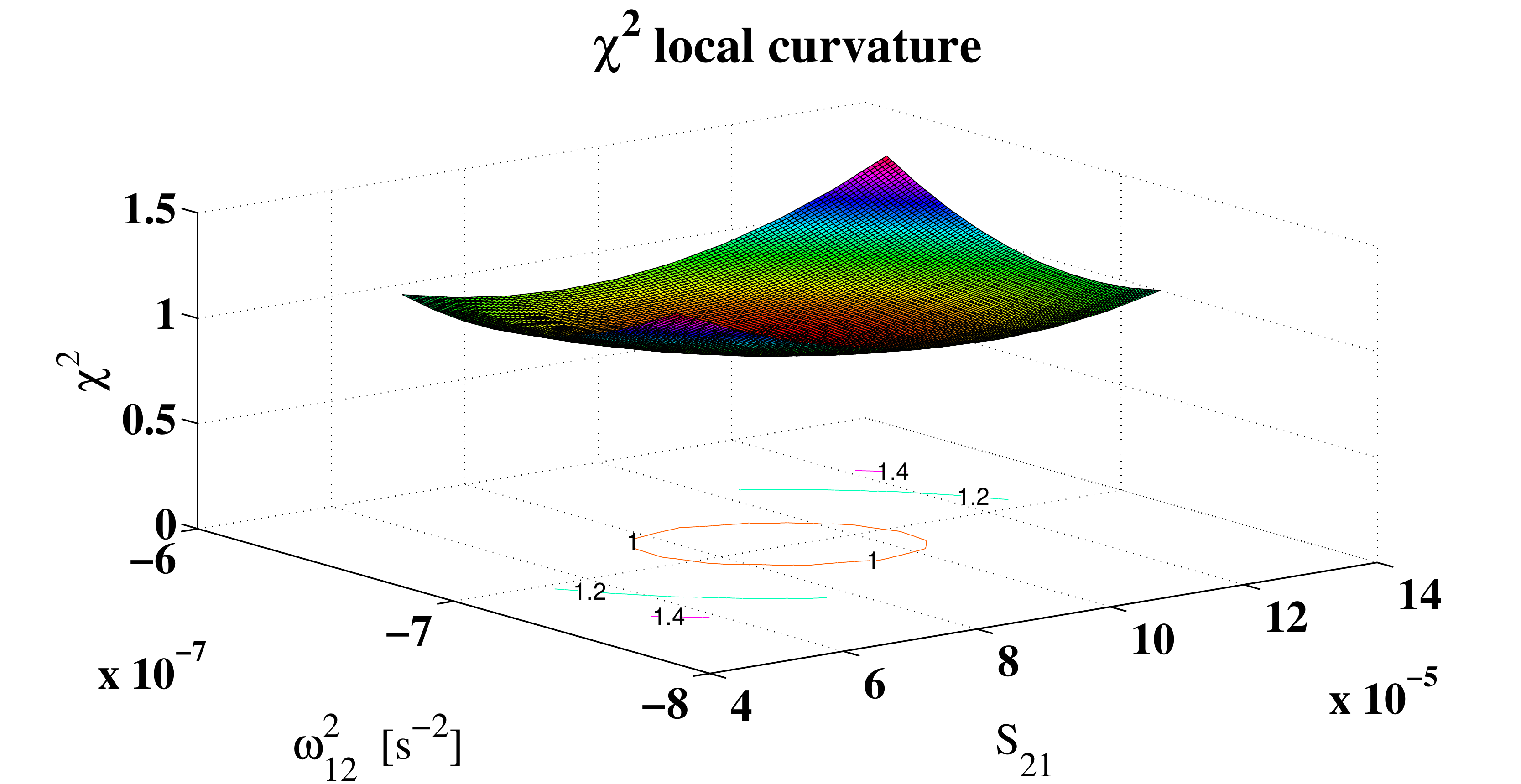}} &
\resizebox{73mm}{!}{\includegraphics{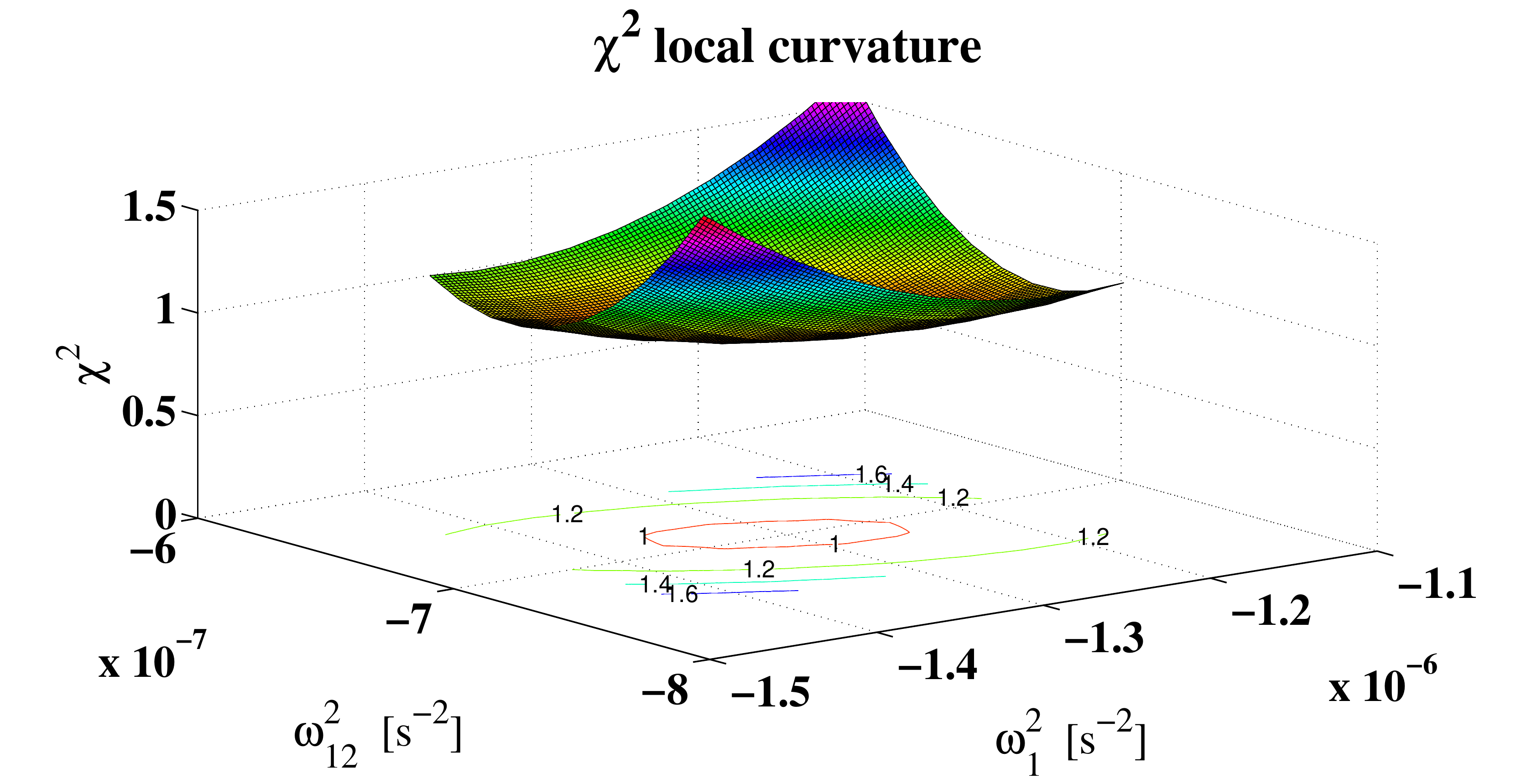}} \\
\hspace{5pt} (c) & \hspace{5pt} (d) \\
\end{tabular}
\caption{\label{fig:montecarlo_chi2_curv}\footnotesize{Fit $\chi^2$ local curvature around the best-fit. The 7-dimensional surface has been projected onto two parameters at a time for some examples. The correlation is responsible for the rotation of the surface. }}
\end{figure*}

The correlation between two parameters is linked to the rotation of the local $\chi^2$ paraboloid around the minimum. To support this statement, we show some plots of the projection of the $7$-dimensional surface onto two parameters at a time, around the best-fit. See Fig.\;\ref{fig:montecarlo_chi2_curv} for some examples. Weakly correlated parameters, like $S_{21}$ and $\omega_1^2$ ($\oforder\;20\%$) (panel (b)) typically have the principal axes of the contour curves aligned with the $x$ and $y$ axes. Highly correlated parameters, like $A_\text{sus}$ and $\omega_1^2$ ($\oforder\;\minus70\%$) (panel (a)) have the principal axes rotated by a non-negligible amount.

\begin{figure}[htb]
\hspace{-10pt}
\includegraphics[width=\columnwidth]{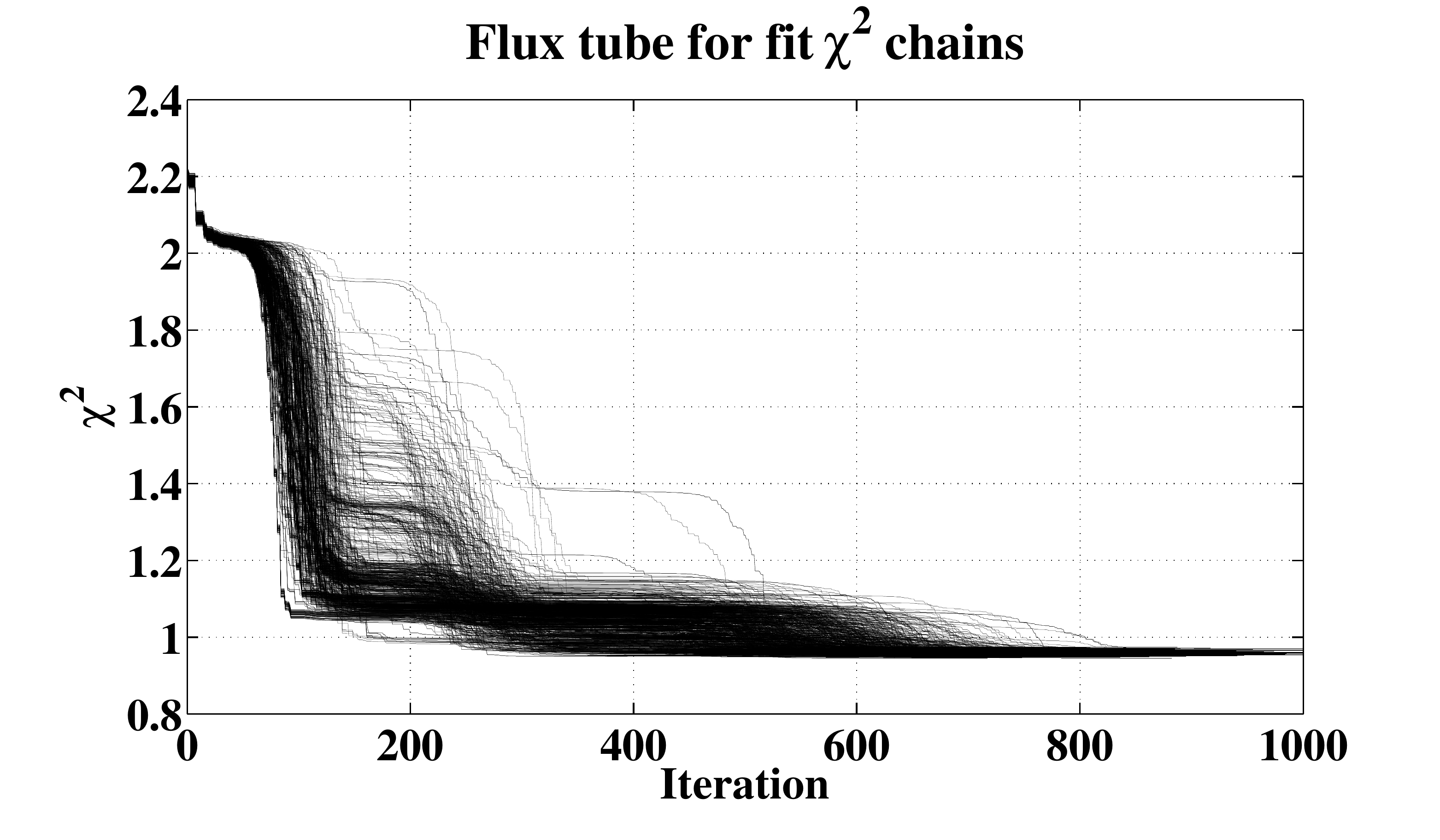}
\caption{\label{fig:montecarlo_chi2_chains}\footnotesize{Monte Carlo fit $\chi^2$ chains. The processes typically last for $\oforder1000$ iterations and stop when either the function or the variable tolerance is below $1\e{\minus4}$.}}
\end{figure}

\begin{figure}[htb]
\hspace{-10pt}
\begin{tabular}{cc}
\includegraphics[width=\columnwidth]{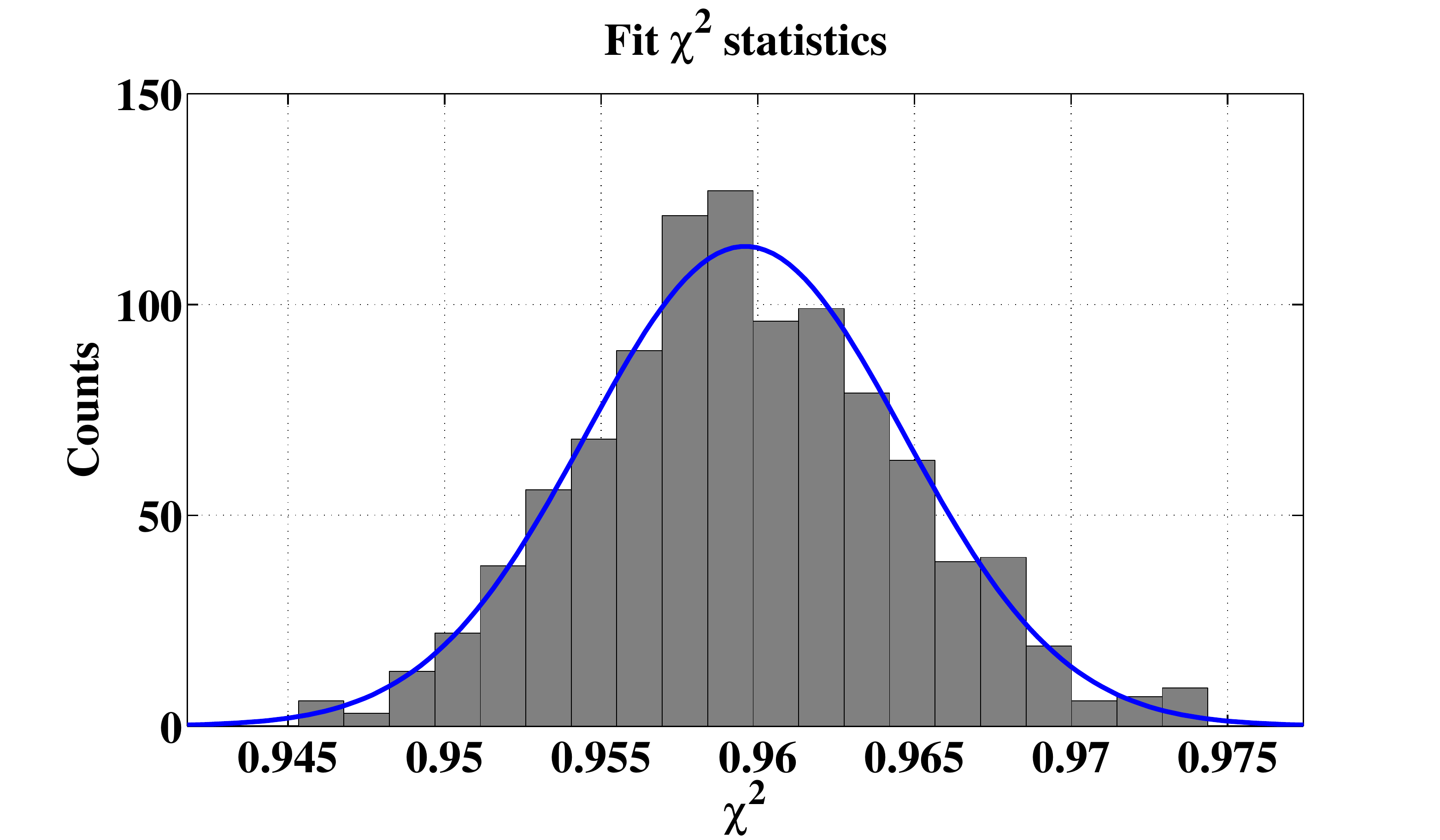} \\
\includegraphics[width=\columnwidth]{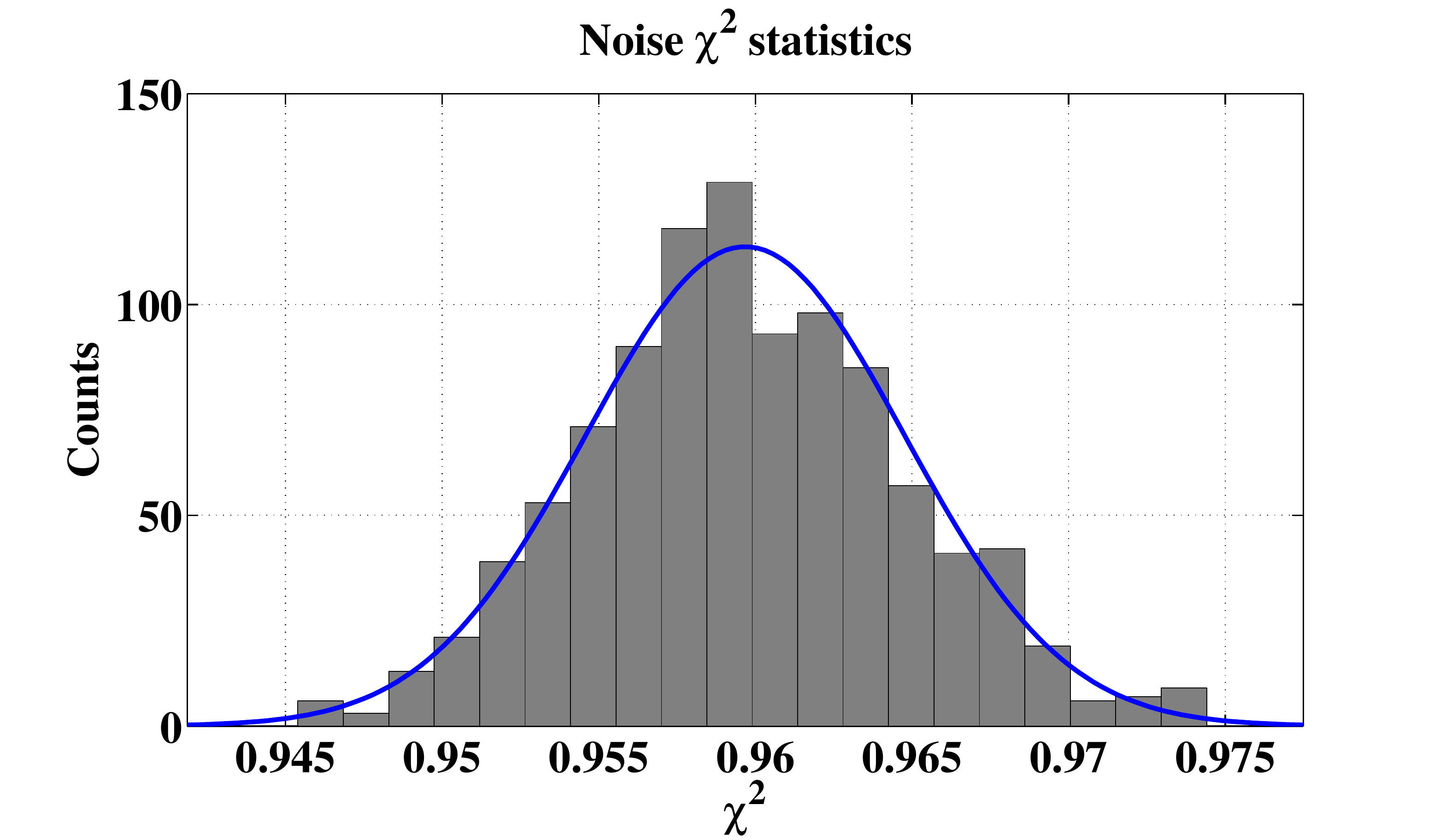}
\end{tabular}
\caption{\label{fig:montecarlo_chi2}\footnotesize{Statistics for $\chi^2$ log-likelihood obtained from the fit and calculated on the noise alone. The agreement between the two shows that the deterministic part of the data is statistically suppressed.}}
\end{figure}

A whole Monte Carlo history of the $\chi^2$ log-likelihood chains is recorded in Fig.\;\ref{fig:montecarlo_chi2_chains}. The scatter of the chains is due to the noise fluctuation among the Monte Carlo iterations. There are clearly some chains that are far away from the accumulation zone: this behavior is completely unexpected as one would think the noise to have little impact on the chains location.
Despite the big scatter, the asymptotic distribution is Gaussian (see Fig.\;\ref{fig:montecarlo_chi2}). The final, and most remarkable check, is the comparison between the fit $\chi^2$ log-likelihood and the one calculated on pure noise data. It is important here to stress that both the fit and the noise $\chi^2$ at a first look showed agreement between each other, but they were both positively skewed. The following facts explain why. In Section \ref{sect:whitening_filters} we have described the practical method to implement the diagonalization of the noise covariance matrix of Section \ref{sect:whitening} with its limitation. This consists in the impossibility of filtering out the lowest frequencies, due to the finiteness of the data stretches from which whitening filters are derived and which causes the skewness. Transparently, the application of a high-pass filter to the data has solved the issue. Finally, the plot provides an important twofold test: on a side, the parameter variances are statistically distributed as the fit $\chi^2$ log-likelihood; on the other, the fit $\chi^2$ log-likelihood is in agreement with the noise $\chi^2$ log-likelihood, showing that the estimation method has suppressed the systematics and estimated the noise statistics with no extra bias.

\subsubsection{Robustness to the initial guess} \label{sect:robust2guess}

We discuss here the robustness of the estimator to the choice of the initial guess.
For this, we start the search from an initial guess for the parameters arbitrarily far away. Physically
this corresponds to a very poor knowledge of the system or to the unlikely (but possible) situation of under-performing actuators and
underestimated couplings between the TMs and the SC. In both cases a precise calibration is needed.

In Table \ref{tab:robust2guess} we summarize the results by comparing the real values,
the initial guess and best-fit. We also indicate the bias (absolute deviation from the real value
in units of standard deviation) for both the best-fit and the initial guess.
We notice that almost all parameters are within $1$ standard deviation from the real
ones, even though the starting guesses are typically at $\oforder10^3$ standard deviations away.
\begin{table}[htb]
\squeezetable
\caption{\label{tab:robust2guess}\footnotesize{Robustness to the initial guess. Initial guess at $\chi^2=1.3\e5$, $\nu=79193$; best-fit at $\chi^2=0.99$. The term in brackets is the error relative to the rightmost digit. In curly brackets the bias (absolute deviation from the real value in units of standard deviation) for each estimate is also shown.}}
\begin{ruledtabular}
\begin{tabular}{l D{.}{.}{1.2} D{.}{.}{1.8} D{.}{.}{1.3} D{.}{.}{2.1} D{e}{\times}{7.-1}}
& \multicolumn{1}{c}{Real} & \multicolumn{2}{c}{Best-fit} & \multicolumn{2}{c}{Guess} \\
\hline
$A_\mathrm{df}$                                     & 0.62      & 0.61994(8)        & \{0.77\}    & 1 & \{4.9$\e{3}$\}    \\
$A_\mathrm{sus}$                                    & 0.6       & 0.599990(8)       & \{1.3\}     & 1 & \{5.1$\e{4}$\}    \\
$S_{21}\,[10^{\minus3}]$                            & \minus1.5 & \minus1.4998(1)   & \{0.55\}    & 0 & \{4.7$\e{3}$\}    \\
$\omega_1^2\,[10^{\minus6}\,\text{s}^{\minus2}]$    & \minus3   & \minus2.9998(2)   & \{1.1\}     & \minus1.3 & \{7.8$\e{3}$\}    \\
$\omega_{12}^2\,[10^{\minus6}\,\text{s}^{\minus2}]$ & \minus2   & \minus2.0000(1)   & \{0.32\}    & \minus0.7 & \{1.0$\e{4}$\}    \\
$\Delta t_1\,[\text{s}]$                            & 0.6       & 0.6013(7)         & \{1.8\}     & 0 & \{8.4$\e{2}$\}    \\
$\Delta t_{12}\,[\text{s}]$                         & 0.4       & 0.398(2)          & \{0.95\}    & 0 & \{2.3$\e{2}$\}    \\
\end{tabular}
\end{ruledtabular}
\end{table}

\begin{figure}[htb]
\hspace{-10pt}
\includegraphics[width=\columnwidth]{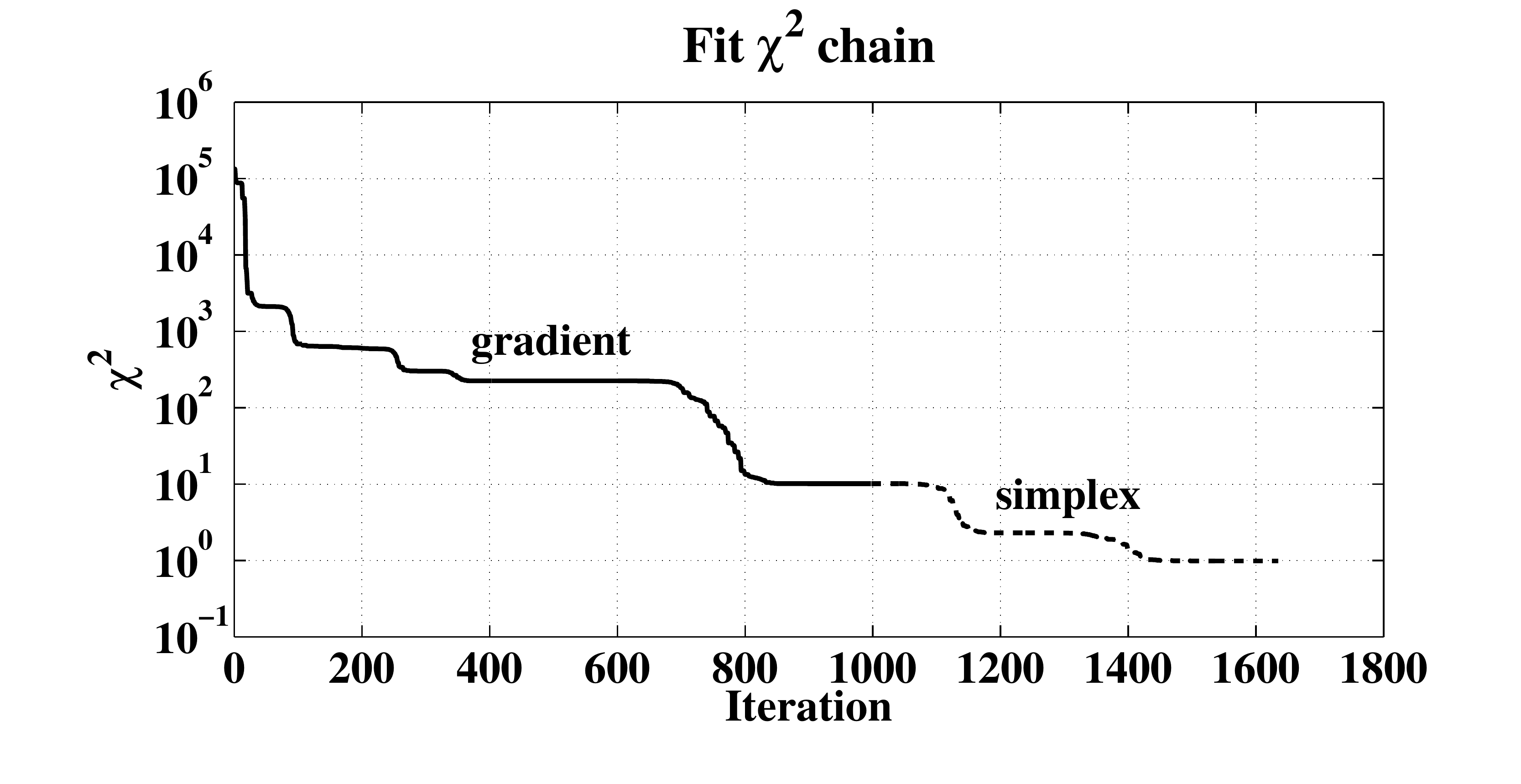}
\caption{\label{fig:robust2guess_chi2}\footnotesize{Fit $\chi^2$ chain that shows the estimation performance from $\oforder1\e{5}$ to the optimum $\oforder1$. The process lasts for 1636 iterations and stops when either the function or the variable tolerance is below $1\e{\minus4}$. A preliminary global gradient search is followed by a local simplex.}}
\end{figure}

In Fig.\;\ref{fig:robust2guess_chi2} we show the performance of the estimation,
showing how the method is able to suppress the $\chi^2$ by many
orders of magnitude, from $1\e{5}$ to $\oforder1$, which is the required optimum, within the given tolerances (set to $1\e{\minus4}$ in both optimization function and variables). Finally, it is clear that we can recover the real values within the confidence level
by decreasing the merit function of many orders of magnitude, while keeping both accuracy and precision.

\begin{figure}[htb]
\hspace{-10pt}
\begin{tabular}{cc}
\includegraphics[width=\columnwidth]{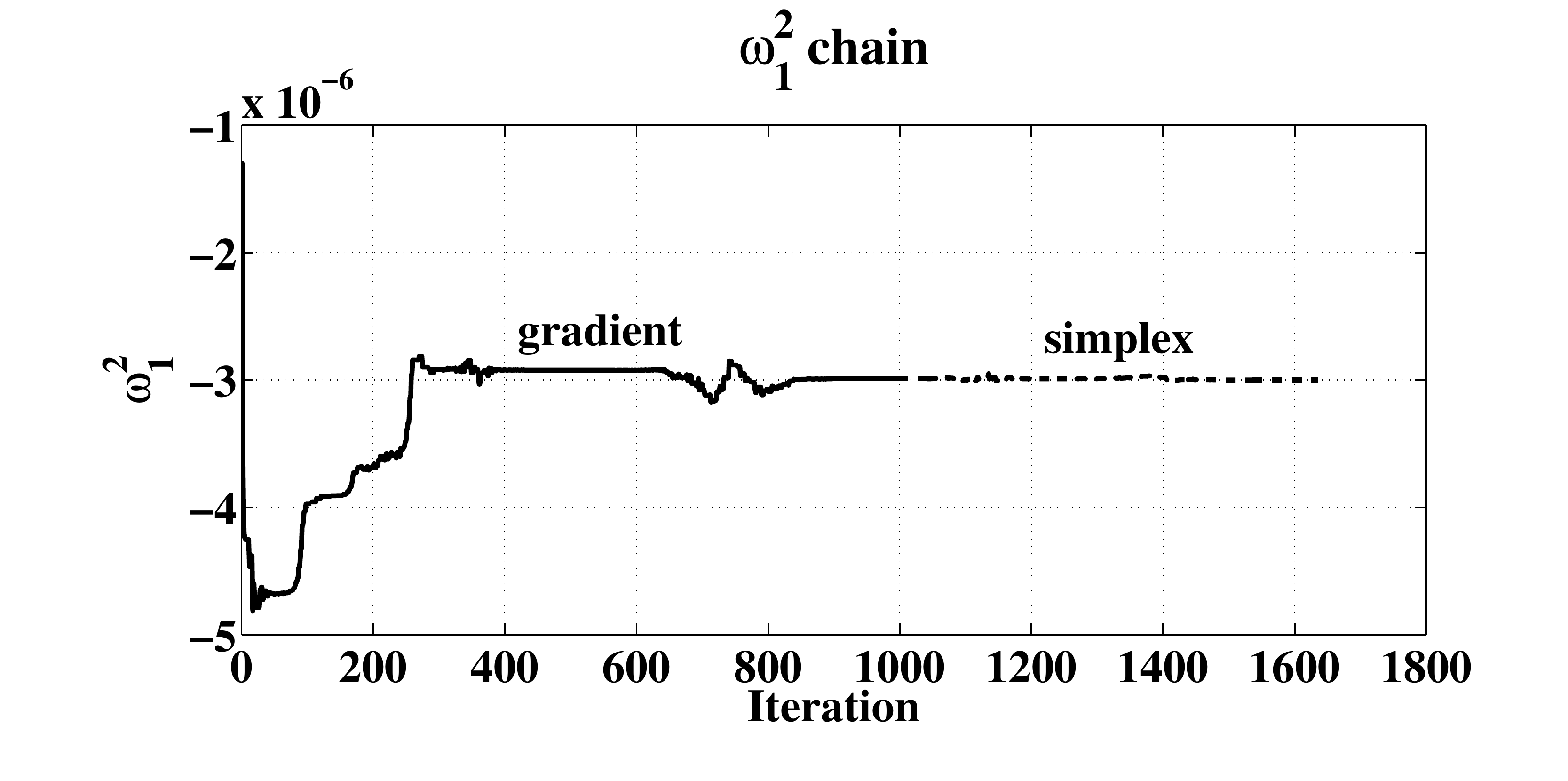} \\
\includegraphics[width=\columnwidth]{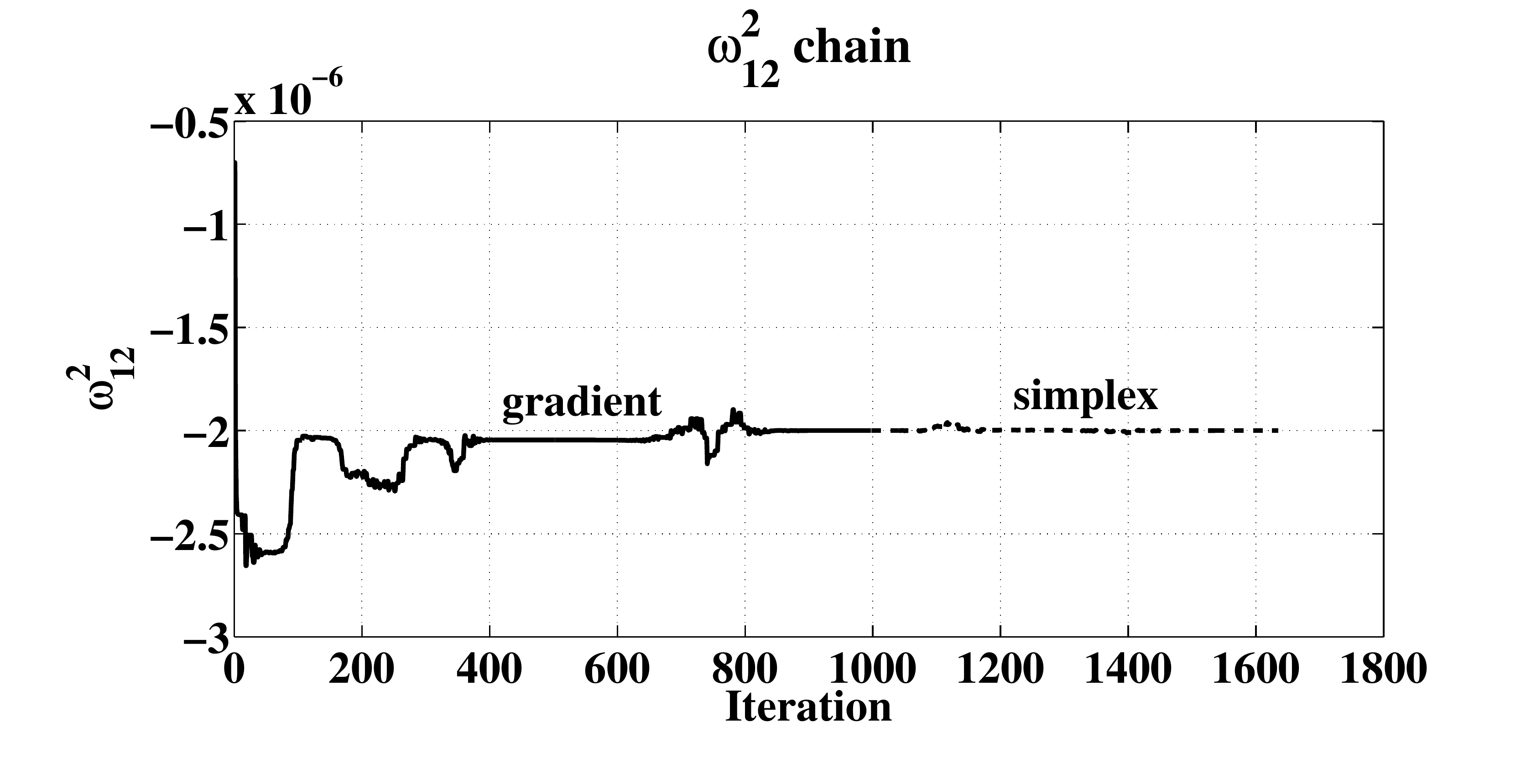}
\end{tabular}
\caption{\label{fig:robust2guess_chain}\footnotesize{Estimation chains for $\omega_1^2$ and $\omega_{12}^2$ during the gradient and simplex searches. The jumps are correlated to those of the $\chi^2$. The saturation to the best-fit is self-evident.}}
\end{figure}

Analogously, in Fig.\;\ref{fig:robust2guess_chain} two examples of estimation chains
(for $\omega_1^2$ and $\omega_{12}^2$) are also reported, showing the correlation with the big jumps of the $\chi^2$ chains and how the parameters saturate to the optima.

\begin{figure*}[htb]
\begin{tabular}{cc}
\resizebox{73mm}{!}{\includegraphics{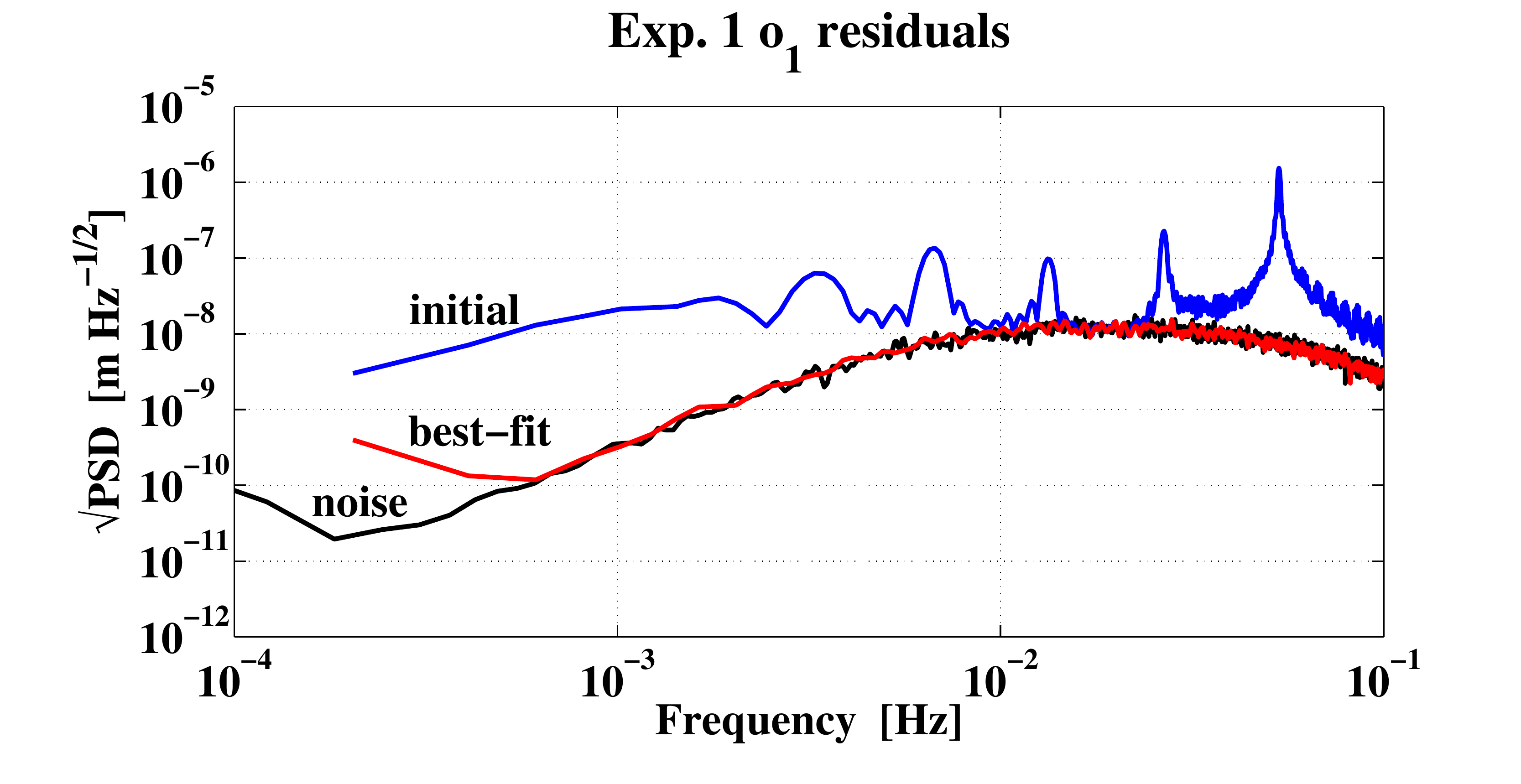}} &
\resizebox{73mm}{!}{\includegraphics{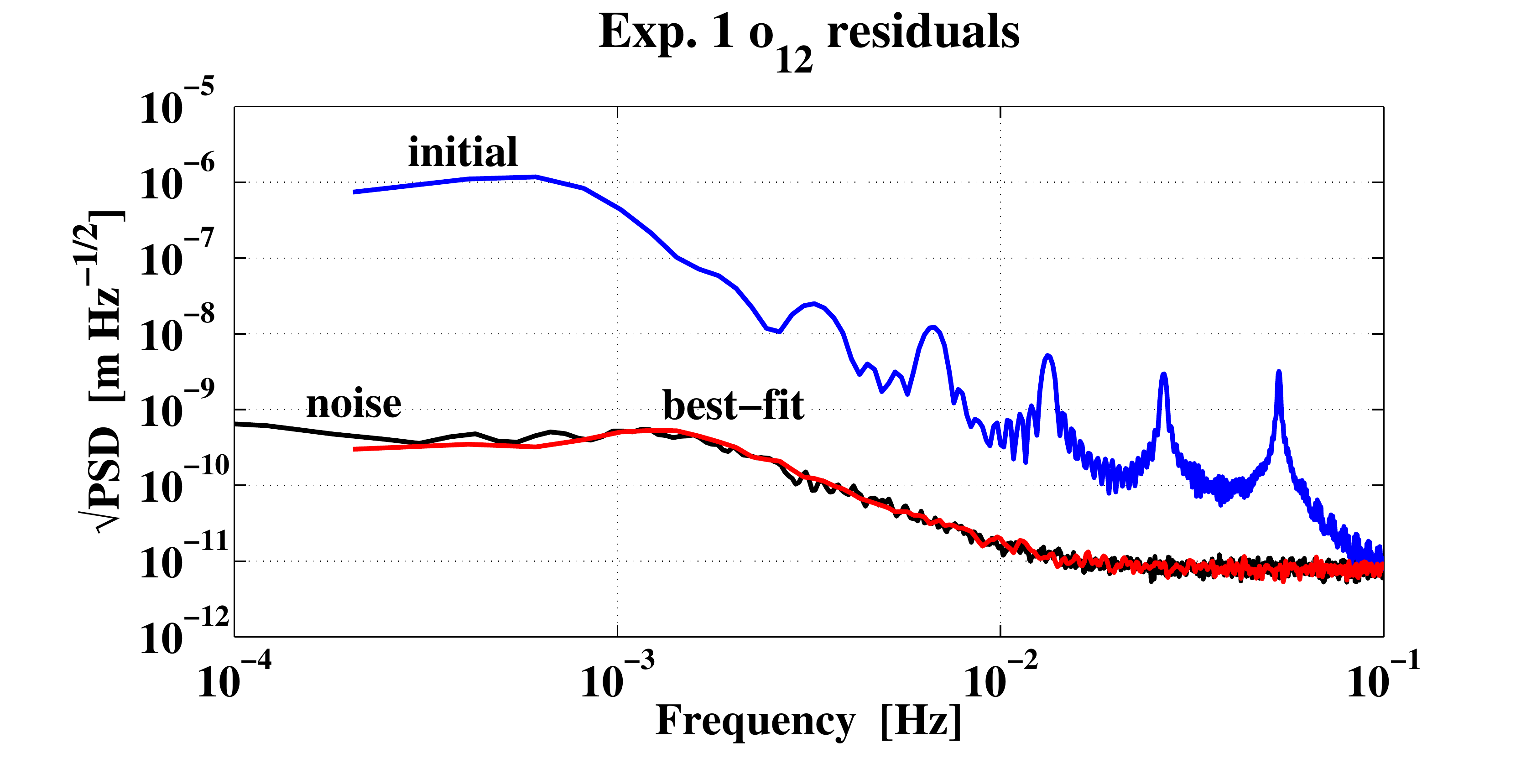}} \\
\hspace{5pt} (a) & \hspace{5pt} (b) \\
\vspace{-5pt} \\
\resizebox{73mm}{!}{\includegraphics{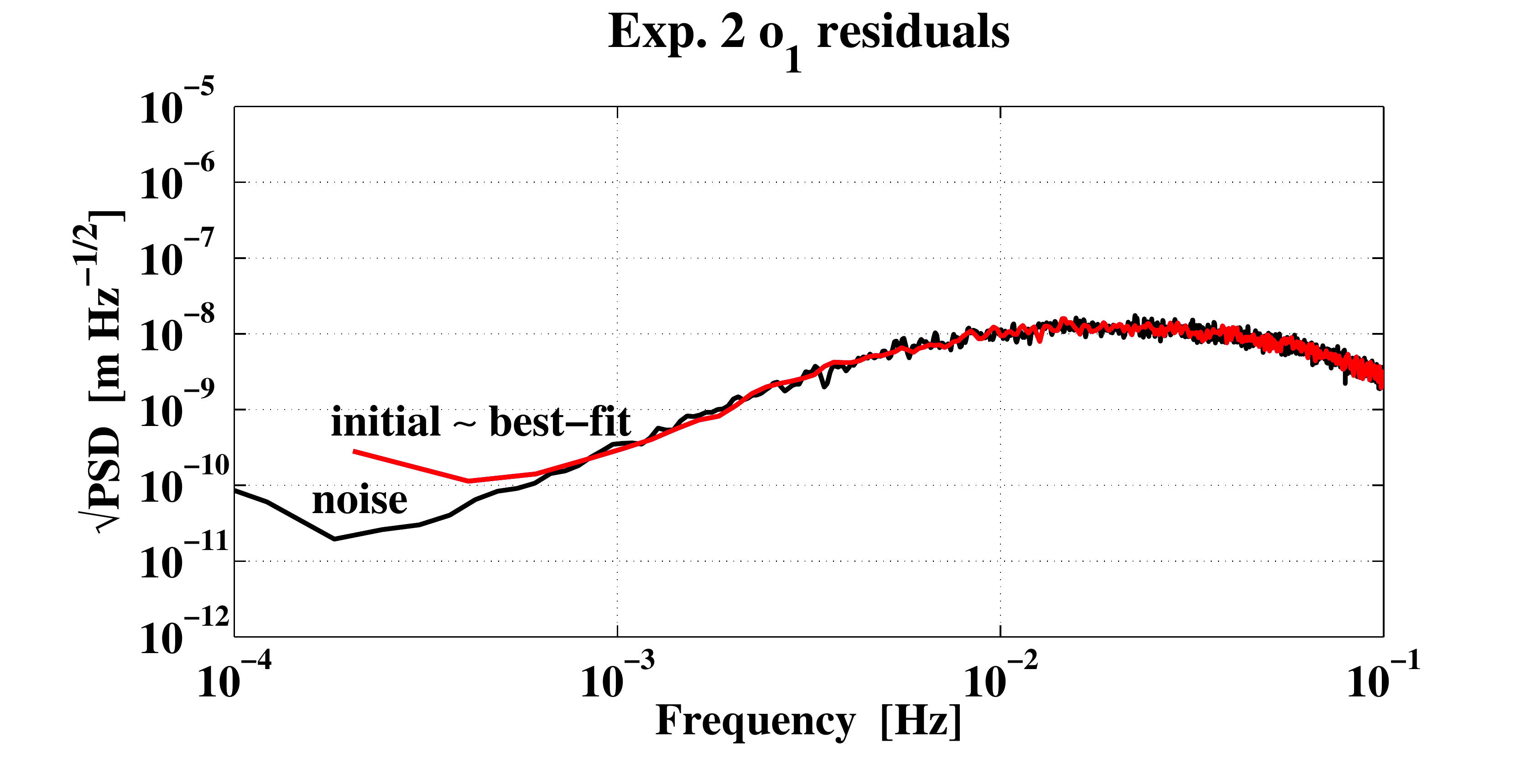}} &
\resizebox{73mm}{!}{\includegraphics{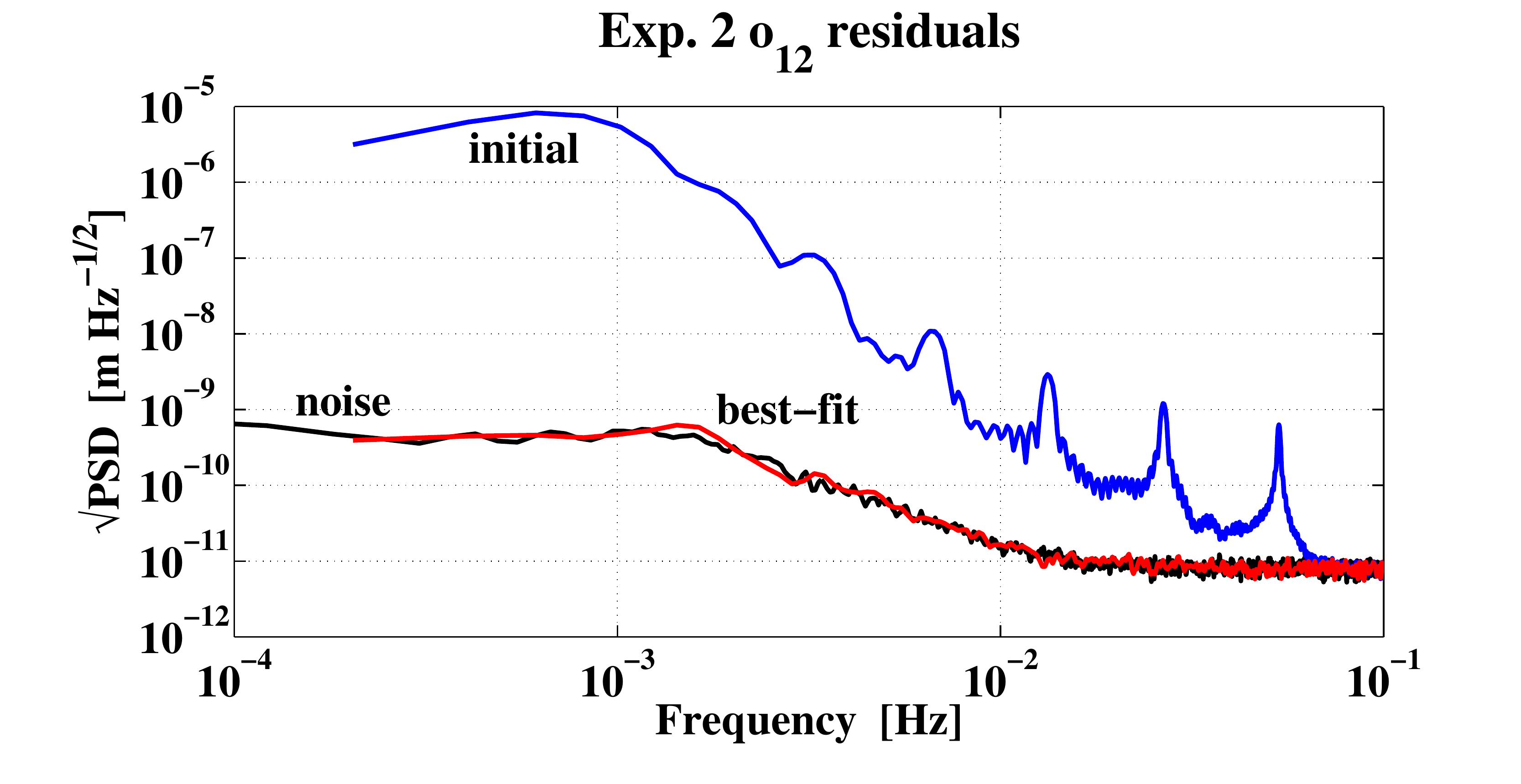}} \\
\hspace{5pt} (c) & \hspace{5pt} (d) \\
\end{tabular}
\caption{\label{fig:robust2guess_residuals}\footnotesize{Analysis of residuals for all simulated experiments and channels. Initial and best-fit residuals are shown and compared to the expected noise shapes. The improvement on the $o_{12}$ channel of both experiments (b) and (d) is of $\oforder 4$ orders of magnitude around 1 mHz. For the $o_1$ channel in the first experiment (a) 3 orders of magnitude; (c) contains no signal. PSDs are computed with the Welch overlap method, 4-sample 92-dB Blackman-Harris window, 16 averages and mean-detrending.}}
\end{figure*}

In the end, the analysis of residuals summarized in Fig.\;\ref{fig:robust2guess_residuals}
demonstrates that it is possible to completely subtract the deterministic
part out of the data and reach the expected noise shapes (estimated from the independent noise run)
for all experiment and channels. The improvement is evident at low frequency:
for the $o_{12}$ channel the residuals are suppressed by $\oforder 4$ orders of magnitude around
\unit[1]{mHz}. The same happens for the $o_1$ channel in the first experiment where
the improvement is of $\oforder 2$ orders of magnitude. Only the $o_1$ channel in the
second experiment contains no signal since the cross-talk from the $o_{12}$
into that channel is negligible.

\subsubsection{Robustness to non-gaussianities} \label{sect:robust2glitches}

This section is devoted to showing the effect of non-gaussianities
in the noise, and how to properly handle them. The main realistic behavior of experimental noise is the presence
of outliers: consequently the sampling distribution of the data may show some prominent
tails. An example of such outliers is the manifestation of glitches, very short
noise transients due to anomalous response in the readout/circuitry.

A standard approach, called \textit{local L-estimate} \cite{press2007}, is to generalize the norm of
residuals in Eq.\;\eqref{eq:chi2}, since it usually overweighs the outliers. The
idea is to properly take care of them by regularizing the usual square of residuals
with other definitions of the norm. As an example, three possible definitions are considered
\begin{equation}
\begin{split}
\langle\mathbf{r}(\mathbf{p})\vert & \mathbf{r}(\mathbf{p})\rangle = \\
&\begin{cases}
\sum_i r_i^2            & \text{mean squared dev.} \\
\sum_i \left|r_i\right| & \text{mean absolute dev.} \\
\sum_i \log(1+r_i^2)    & \text{mean logarithmic dev.}
\end{cases}~,
\end{split}
\end{equation}
corresponding to the Gaussian, log-normal and Lorentzian distribution, respectively.
The subscript $i$ counts the data and the channels/experiments.

The method can be successfully applied to data with glitches.
Noise glitches are unpredictable high frequency noise transients mostly
due to failures in the circuitry. Such outliers usually fall well beyond $3$ standard
deviations and produce an excess at the tails of the statistic. Since the output of the interferometer might be subject to similar phenomena, we simulate a realistic experiment containing glitches and model such transients as Sine-Gaussian (SG)
functions
\begin{equation}
n_\text{GL}(t) = a\sin(2\pi f_0(t-t_0))\exp \left(\frac{\minus(t-t_0)^2}{\tau^2}\right)~,
\end{equation}
where the SG parameters span a wide (uniformly distributed) range of values. In particular,
the SG frequency, $f_0$, covers the whole bandwidth $\unit[(10^{\minus4}-0.45)]{Hz}$;
the injection time, $t_0$, is distributed all along the whole time-series;
the characteristic time, $\tau$, which gives the typical duration of the pulse
is $\unit[(1-2)]{s}$; the amplitude, $a$, is defined in terms of the number of
the original noise standard deviations and falls outside the Gaussian statistic by
$(3-20)\;\sigma_\text{n}$. Moreover, we fix the number of SG injections as
a fractional part of the whole data series: we conventionally choose $f_\text{SG}=N_\text{SG}/N_\text{data}=1\%$,
since higher values are very unlikely. Notice that this value represents only the number of injections: the actual fraction of corrupted data will be of the order of
$3\text{E}[\tau]f_\text{SG}\simeq5\%$.

\begin{figure}[htb]
\includegraphics[width=\columnwidth]{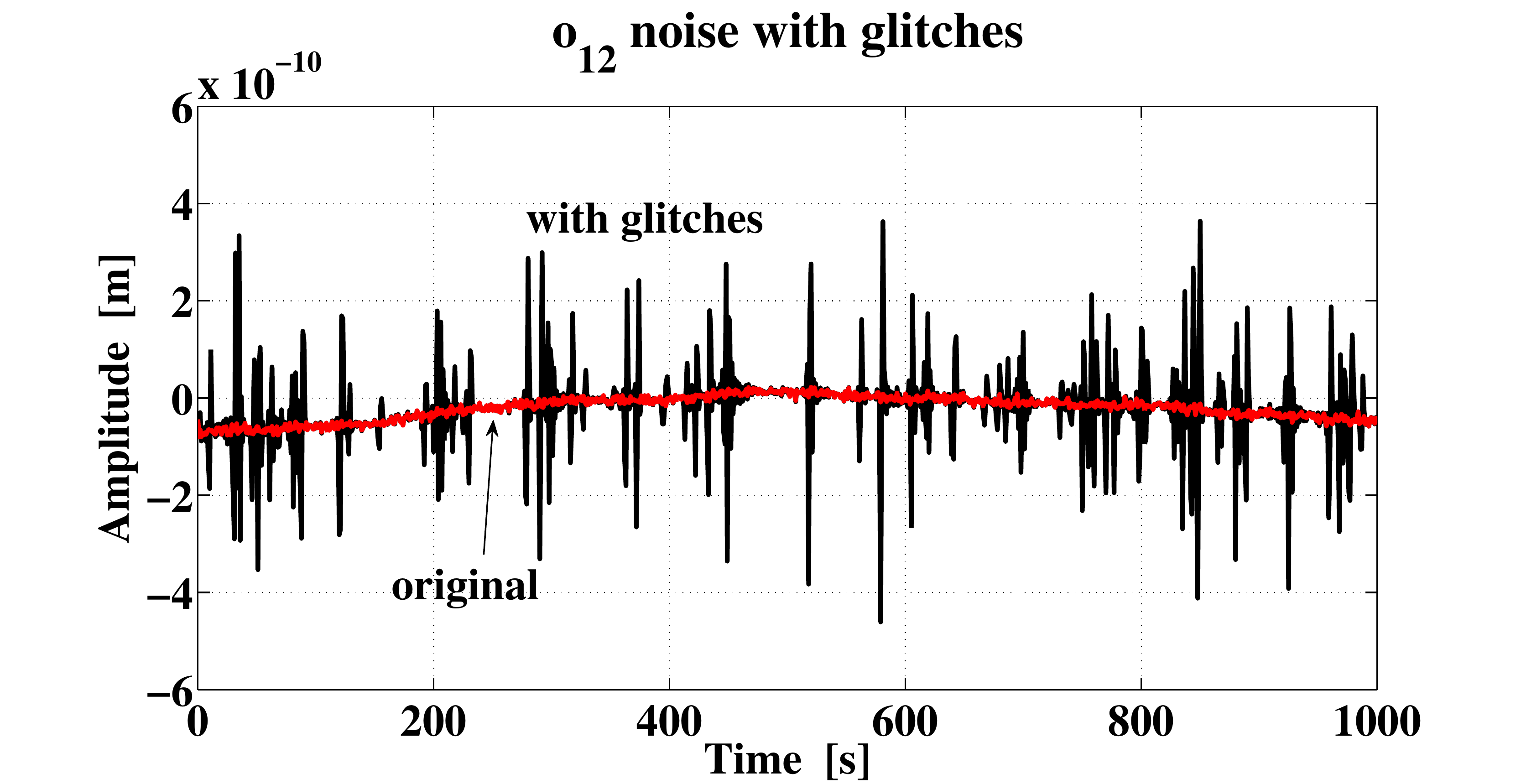}
\caption{\label{fig:robust2glitches_glitches}\footnotesize{Original and glitchy noise for the $o_{12}$ channel. The level of data corruption is evident.}}
\end{figure}

Glitchy noise is readily produced by coloring a white, zero-mean, unitary standard
deviation input time-series, corrupted by random
injections of SGs. See Fig.\;\ref{fig:robust2glitches_glitches} for an example
of injection of glitches into the differential channel. The simulated noise is
now what the experiment would give us during the real-life mission.

\begin{figure}[htb]
\includegraphics[width=\columnwidth]{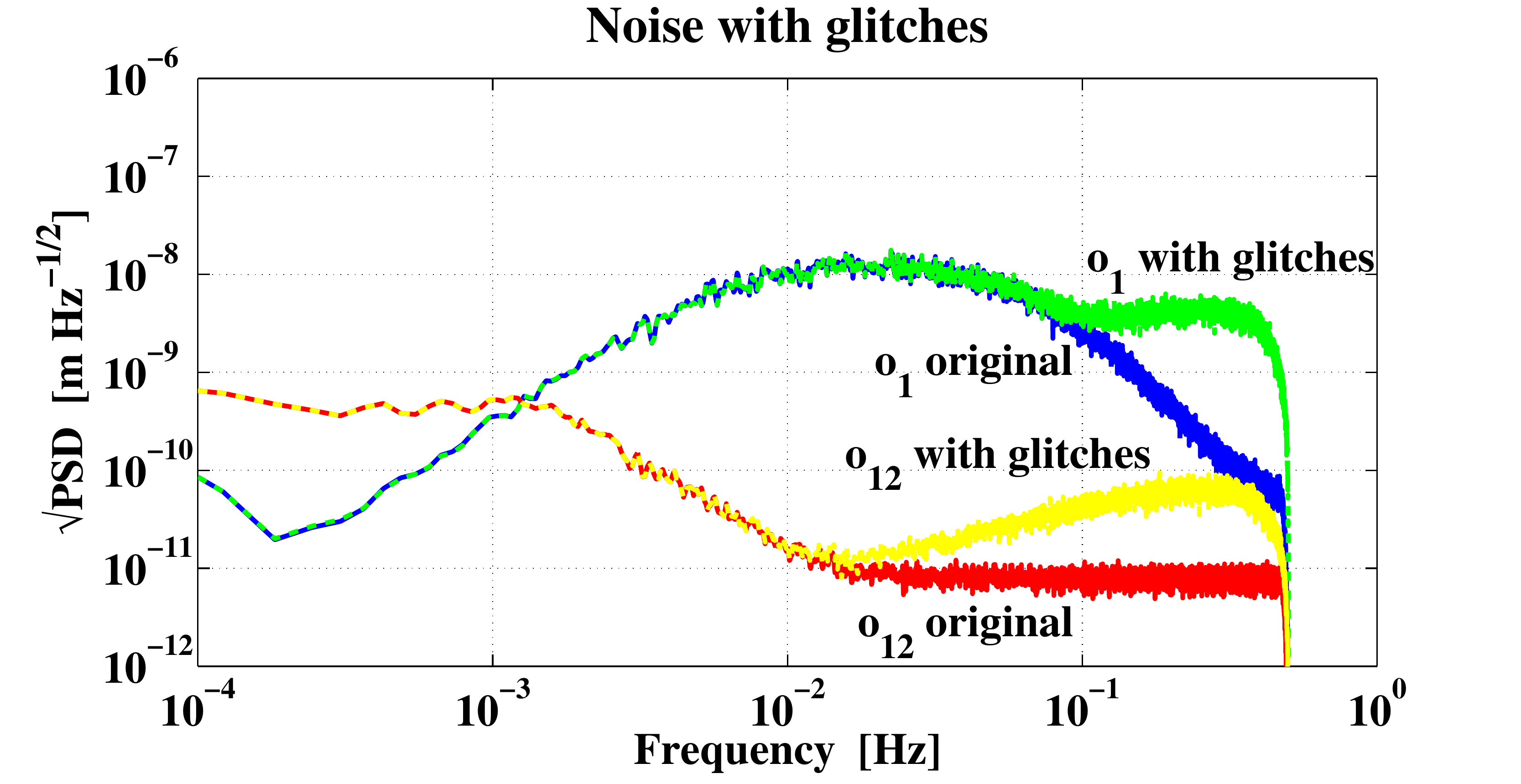}
\caption{\label{fig:robust2glitches_noise}\footnotesize{Spectra for simulated glitchy noise compared to the original for $o_1$ and $o_{12}$ channels. The high frequency bump around 0.2 Hz is evident. PSDs are computed with the Welch overlap method, 4-sample 92-dB Blackman-Harris window, 16 averages and mean-detrending.}}
\end{figure}

The effect of these glitches is that the PSD of the simulated noise scales linearly
with the frequency, up to $\unit[4\e{\minus9}]{m\,Hz^{\minus1/2}}$ and $\unit[6\e{\minus11}]{m^2\,Hz^{\minus1}}$
around \unit[0.2]{Hz} for the first and differential
channel, respectively. This excess noise sums up to the original one and affects its high frequency components (see Fig.\;\ref{fig:robust2glitches_noise}). Obviously, the new statistic contains an excess at the tails.
For example, the $o_1$ channel has a kurtosis of $19$, compared to the original
one of $\minus9\e{\minus3}$. No significant difference in skewness is detected since
the statistic does not loose symmetry after the SG injections.

In order to perform a realistic parameter estimation, whitening filters are derived
from the glitchy noise stretches with the same procedure described in Section
\ref{sect:whitening_filters}. However, since the whitening works only in the stationary
regime and the glitches are highly non-stationary, it is practically impossible
to filter them out of the data.

Table \ref{tab:robust2glitches} shows the results of the parameter estimation using
the three norm definitions introduced above.
\begin{table*}[htb]
\squeezetable
\caption{\label{tab:robust2glitches}\footnotesize{Robustness to glitches. We compare the norm for squared, absolute and logarithmic deviation with $\nu=79193$. The term in brackets is the error relative to the rightmost digit. In curly brackets the bias for each estimate is also shown.}}
\begin{ruledtabular}
\begin{tabular}{l D{.}{.}{3.4} D{.}{.}{1.5} l D{.}{.}{1.5} l D{.}{.}{1.5} l D{.}{.}{3.4}}
& \multicolumn{1}{c}{\multirow{2}{*}{Real}} & \multicolumn{2}{c}{Best-fit} & \multicolumn{2}{c}{Best-fit} & \multicolumn{2}{c}{Best-fit} & \multicolumn{1}{c}{\multirow{2}{*}{Guess}} \\
& & \multicolumn{2}{c}{(mean sq.\;dev.)} & \multicolumn{2}{c}{(mean abs.\;dev.)} &  \multicolumn{2}{c}{(mean log.\;dev.)} \\
\hline
Norm                                   &           & \multicolumn{2}{c}{10}                                & \multicolumn{2}{c}{2.1}     & \multicolumn{2}{c}{0.95} \\
\hline
$A_\mathrm{df}$ & 1.01 & 1.011(3) & \{0.29\} & 1.010(1) & \{0.23\} & 1.0109(8) & \{1.2\} & 1 \\
$A_\mathrm{sus}$ & 0.99 & 0.99000(5) & \{0.035\} & 0.98959(2) & \{20\} & 0.99001(1) & \{0.99\} & 1 \\
$S_{21}\,[10^{\minus4}]$ & 1.1 & 1.10(2) & \{0.074\} & 1.113(7) & \{1.8\} & 1.116(5) & \{3.4\} & 0 \\
$\omega_1^2\,[10^{\minus6}\,\text{s}^{\minus2}]$ & \minus1.32 & \minus1.320(1) & \{0.061\} & \minus1.3188(6) & \{2.0\} & \minus1.3192(4) & \{2.0\} & \minus1.3 \\
$\omega_{12}^2\,[10^{\minus6}\,\text{s}^{\minus2}]$ & \minus0.68 & \minus0.6798(7) & \{0.29\} & \minus0.68000(3) & \{0.011\} & \minus0.6804(2) & \{1.8\} & \minus0.7 \\
$\Delta t_1\,[\text{s}]$ & 0.1 & 0.100(3) & \{0.045\} & 0.090(1) & \{8.3\} & 0.1007(8) & \{0.90\} & 0 \\
$\Delta t_{12}\,[\text{s}]$ & 0.1 & 0.098(5) & \{0.36\} & \minus0.0290(2) & \{58\} & 0.098(2) & \{1.2\} & 0 \\
\end{tabular}
\end{ruledtabular}
\end{table*}

The most conservative least square estimator
provides overestimated errors since they scale as $\oforder\sqrt{\chi^2}$.
The absolute and logarithmic deviation provide better statistics and lower errors,
but the first gives biased estimates of $A_\mathrm{sus}$, $\Delta t_1$ and
$\Delta t_{12}$ and the second a slightly biased estimate of $S_{21}$. The analysis
of residuals (not shown here) demonstrates that the methods are in agreement
with each other and the noise shapes. From the performance point of view, these
estimators are $30\%$ and $9\%$ faster than the Gaussian (mean squared deviation). As we can see,
there is no absolute rule we can apply when dealing with glitches. However, from the difference
between the methods we can infer the sensitivity of each single parameter
to glitches. For example, we can assume the ratio between the biases as the criterion for comparing two methods. The ratio tends to one if that parameter is not sensitive to glitches; otherwise, tends to a very small or big number. The comparison between the mean squared
deviation and the mean logarithmic deviation gives that $S_{21}$ is the most
sensitive parameter, whilst $\Delta t_{12}$ the least. Therefore, starting
from the fact that the methods give the same result for purely Gaussian noise,
our proposed recipe is the following:
\begin{enumerate}
\item apply the conservative approach (the ordinary mean squared deviation) directly to corrupted time series and try with different estimators (mean absolute deviation, mean logarithmic deviation, etc.);
\item start to remove some outliers giving them negligible weight;
\item redo the analysis with all estimators;
\item check for convergence and agreement between the estimators.
\end{enumerate}
The whole process is actually a reweighing analysis that provides robust
uncertainties and finally removes the outliers in a step-by-step smooth readjustment.
Even though it would be possible in principle to clean up the data
just before the estimation, in that case the results would likely be dependent on
the statistical criterion. The main advantage of our recipe is its robustness
and the fact that the data polishing is strictly step-by-step and model independent.

\subsection{Analysis of operational exercises} \label{sect:stoc}

The estimation of parameters has been a pivotal subject during many data analysis sessions in view of the LPF mission. The core
experiments in dynamics are purposefully described in the LPF Experiment Master Plan and envision system identification methods of the kind described in Section \ref{sect:dynamics}.

With the aim of validating the data analysis effort and the conversion of the science strategies into tele-commands and on-board instructions (Payload Operations Requests), several
extended data analysis exercises were called in the past 2 years. At first they took the form of Mock Data Challenges with two parties involved: data generation and data analysis. Rapidly the need of
closing and testing the chain from science to payload demanded the evolution of these data analysis sessions into real Operations Exercises with a few team leaders in co-location and the Principal Investigators' teams on call, to mimic the real mission time.

In absence of the real LTP a simulator was used to generate the data. The Offline Simulation Environment (OSE, sometimes called Drag Free and Attitude Control System end-to-end simulator) provided by ASTRIUM \cite{astrium} is a piece of software embedding the same dynamics and most of the software that will run on the on-board computer of LPF, with the advantage of being coded in a set of C/C++ modules called by a MATLAB$^\circledR$ and Simulink$^\circledR$ \cite{simulink} engine, therefore easier to handle than the real operational machinery. The OSE contains detailed State-Space models of the most relevant disturbances and allows for their fine-tuning and (de)activation; it simulates the dynamics, the capacitive and thrusters dispatching algorithms, handles possible couplings between different degrees of freedom as well as the controller choices and decoupling strategy proper of each driving mode of LPF. The OSE is thus equivalent to an-silico LPF laboratory, crucial for characterizing the behavior of the signals and telemetry on-board.

The first phase of validation of the system identification experiments culminated with the sixth Operational Exercise, where the concepts explained in
Sections \ref{sect:maximum_likelihood} and \ref{sect:data_analysis} insofar were applied to the OSE telemetry. The Exercise targeted the estimation of
parameters using a linear fit with Singular Value Decomposition, a non-linear fit (insofar described) and a Markov-chain Monte-Carlo method.

In practice, for the linear and non-linear fit the following recipe was followed:
\begin{enumerate}
\item identification of the time section where the signals had been injected;
\item creation and training of whitening filters from the $60000\,\unit{s}$ of noise data before the injection;
\item whitening
\item creation of numerical template for the injected signals and the dynamics
\item fitting of parameters
\item subtraction of the so modelled signal from the data
\item estimation of the residual acceleration noise using the estimated parameters.
\end{enumerate}

The parameters (see Eq. \eqref{eq:eom_1D}) were not fitted directly for numerical stability issues: it is more efficient to fit the discrepancies with regard to the nominal values, guessed from common sense or obtained from ground-testing experiments. See Table \ref{tab:fitparams} for a full list. Physical parameters can be recovered inverting the parameters expansions and the fit uncertainty can be propagated according to the uncertainty propagation theory, e.g.
\begin{equation}
\begin{split}
\var\left[A_\text{df}\right] =& \var\left[\delta A_\text{df}\right] \\
\var\left[\omega_2^2\right] =& (1.9\e{\minus6})^2 \var\left[\delta \omega_1^2\right] + (1\e{\minus7})^2 \var\left[\delta \omega_{12}^2\right] \\
&+ 2\times(1.9\e{\minus6})\times (1\e{\minus7})\cov\left[\delta \omega_1^2, \delta \omega_{12}^2\right] \\
\ldots & \\
\end{split}
\end{equation}

\begin{table*}[htb]
\squeezetable
\caption{\label{tab:fitparams}\footnotesize{Physical parameters for LPF and their expansion around nominal (or ground tested) values to obtain the parameters used for fitting. The true value was set in the OSE configuration, while the error was computed as the inverse of the Fisher information matrix. Initial values for the fit were set to $0$ for all fitting parameters. The rationale behind the choices is briefly described.}}
\begin{ruledtabular}
\begin{tabular}{p{4.5cm}p{2cm}p{1.4cm}p{1.4cm}p{1cm}l}
Physical parameter and functional expansion & True value (simulator) & Expected error (Fisher) & Fit parameter & Initial guess & Rationale \\
\hline
$A_\text{df} \simeq 1 + \delta A_\text{df}$ & $1$ & $4\e{\minus 4}$ & $ \delta A_\text{df}$ & $0$ & ASTRIUM, controller specification.\\
$A_\text{sus} \simeq 1 + \delta A_\text{sus}$ & $1$ & $2\e{\minus 5}$ & $ \delta A_\text{sus}$ & $0$ & ASTRIUM, controller specification.\\
$S_{11} \simeq 1 + \delta S_{11}$ & $1$ & $-$ & $-$ & $0$ & Ground-testing measure: $0.0000(1)$.\\
$S_{12} \simeq \delta S_{12}$ & $0$ & $-$ & $-$ & $0$ & Ground-testing measure: $0.000(1)$.\\
$S_{21} \simeq \delta S_{21}$ & $1$ & $3\e{\minus 7}$ & $\delta S_{21}$ & $0$ & Simmetry with respect to $\delta S_{21}$.\\
$S_{22} \simeq 1 + \delta S_{22}$ & $0$ & $-$ & $-$ & $0$ & Ground-testing measure: $0.0000(1)$.\\
$\omega_1^2 \simeq 1.9\e{\minus6} (1+\delta \omega_1^2) \,[\text{s}^{\minus2}]$ & $\minus1.307\e{\minus 6}$ & $1\e{\minus 9}$ & $\delta \omega_1^2$ & $0$  & Ground testing measures, worst-case.\\
$\omega_{12}^2 \simeq 1\e{\minus7} (1+\delta \omega_{12}^2) \,[\text{s}^{\minus2}]$ & $\minus6.92\e{\minus 7}$ & $5\e{\minus 10}$ & $\delta \omega_{12}^2$ & $0$ & Ground testing measures, worst-case.\\
$\omega_2^2 = \omega_1^2 + \omega_{12}^2 \,[\text{s}^{\minus2}]$ & $\minus1.999\e{\minus 6}$ & $-$ & $-$ & $-$ & Derived parameter. \\
$\Delta t_1 \,[\text{s}]$ & $\minus 0.2$ & $-$ & $\Delta t_1$ & $0$ & Simulator setup.\\
$\Delta t_2 \,[\text{s}]$ & $\minus 0.2$ & $-$ & $\Delta t_2$ & $0$ & Simulator setup.\\
\end{tabular}
\end{ruledtabular}
\end{table*}

The expected errors on parameters were estimated as optimal errors by inverting the Fisher information matrix: the error is therefore function of the input signals, the model used to describe the experiments and the noise.

The techniques elucidated in Section \ref{sect:param_est} were employed with a linear and non-linear fitting scheme. During the Exercise
a Bayesian method using Markov chains was used too, a detailed description of which is to be found in \cite{nofrarias2010}. The agreement between the methods is very good and their comparison will be discussed elsewhere.

As a matter of fact, the application of the whitening and fitting techniques to an operational scenario proved to be an excellent blind test in guessing the true values of the parameters in Table \ref{tab:fitparams}.
The values obtained in the fit matched the true values within a few standard deviations but showed some extra systematics and overshooting. The issue was highlighted by the analysis of residuals which showed the full matching of the noise, with the exception of a little mismatch at high frequency. The OSE is in fact a Multi-Input-Multi-Output simulator whose internal structures reflect a more realistic and cross-connected model of LPF, not necessarily embedding our reduced matrices (see Eq.\;\eqref{eq:eom_1D}) without extra non-diagonal or cross-controller contamination terms. In particular, the OSE engine dispatching forces and torques to the (virtual) LPF actuators is designed around a decoupling strategy that --- in spite of the name --- truly couples the dynamics of several
degrees of freedom to attain a cleaner and steadier readout in the $o_{12}$ channel. More experiments and an enhanced model shall shine some light on the matter and improve our understanding.

\subsection{Estimation of residual force noise} \label{sect:force_noise}

This final part justifies the importance of the method proposed insofar. As said, the main scientific target of LPF is the characterization of
the TM to TM laser link to allow for detection of GWs in LISA.
In order to fulfill this plan, the required level of differential force noise must be achieved.
If we are pessimistic and assume our knowledge of the key
system parameters is poor, or equivalently we have an under-performing system (as in Section \ref{sect:robust2guess}), we show here
that without a precise calibration of the system dramatic systematic errors might
arise.

In the analysis of operation exercises we have stressed the importance of parameter estimation in the evaluation of the residual force noise.

The solution of the system equations \eqref{eq:eom_matrix1}-\eqref{eq:eom_matrix3} allows for an exact computation of the out-of-loop residual force noise: in fact, by applying the operator of Eq.\;\eqref{eq:ifo2acc} on noisy interferometric data the control and the all known forces can be isolated and subtracted. We have simulated the LPF noise by means of a parametric projection of the individual noise sources to the interferometric readout. By feeding the noise sum to the interferometer model, the readouts were simulated in turn. Such computation has been performed assuming the real parameters (true noise), the initial guess (without identification) and the best-fit (with identification).

\begin{figure*}[htb]
\resizebox{15cm}{!}{\includegraphics{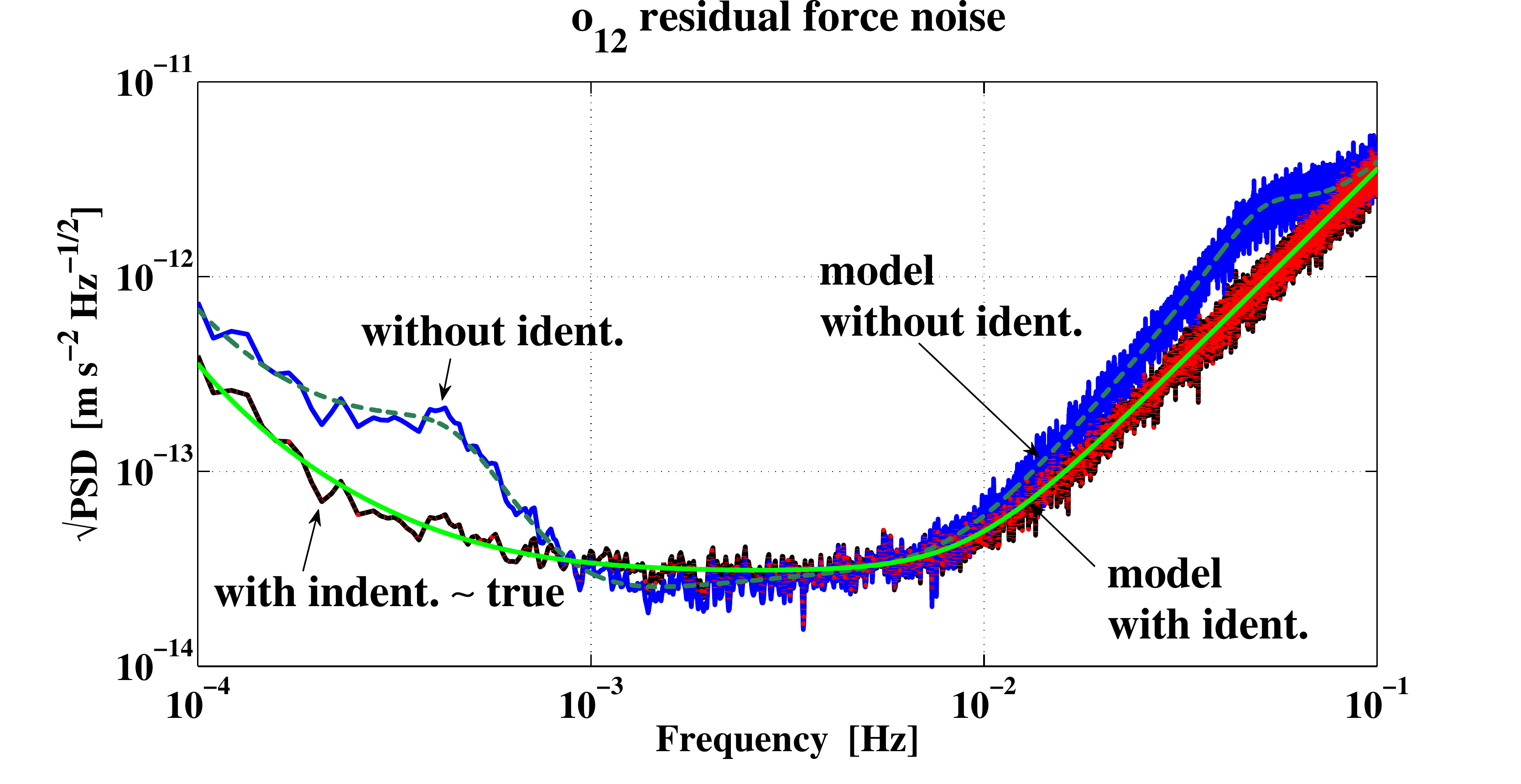}}
\caption{\label{fig:force_noise}\footnotesize{Total residual force noise (per unit mass) numerically estimated on synthetic data and compared to theoretical noise models obtained from a projection of fundamental noise sources. We show different results for numerical estimates and models corresponding to the assumption of the initial parameters (without identification), the best-fit (with identification) and the true parameters (true). The agreement between the models and the estimated noise is clear. The parameter estimation method described in this paper allows the full reconstruction of the true noise shapes and an improvement of a factor 4 at low frequency. PSDs are computed with the Welch overlap method, 4-sample 92-dB Blackman-Harris window, 16 averages and mean-detrending.}}
\end{figure*}

The result of such a computation on a very long noise run, $\oforder 6$ days, is reported in Fig.\;\ref{fig:force_noise}.
We show there the PSD of the residual force noise per unit mass in the $o_{12}$ channel. As is clear from the plot, the best-fit curve reconstructs the true noise shape. The curves can be compared with a theoretical projection of the interferometer noise model to the residual force noise, following the prescription
\begin{equation}
\mathbf{S}_{\text{n},f}(\omega,\mathbf{p}) = \mathbf{H}_{o \rightarrow f}^\dag(\omega,\mathbf{p})\cdot \mathbf{S}_{\text{n},o}(\omega,\mathbf{p}_\text{true})\cdot\mathbf{H}_{o \rightarrow f}(\omega,\mathbf{p})~,
\end{equation}
where $\mathbf{H}_{o \rightarrow f}(\omega,\mathbf{p})$ is the matrix of transfer function models of Eq.\;\eqref{eq:ifo2acc} and $\mathbf{S}_{\text{n},o}(\omega,\mathbf{p}_\text{true})$ is the cross-spectral density matrix of the interferometer readout evaluated at the true parameter values. The plot shows the noise models obtained maintaining $\mathbf{S}_{\text{n},o}(\omega,\mathbf{p}_\text{true})$ fixed and evaluating $\mathbf{H}_{o \rightarrow f}(\omega,\mathbf{p})$ at the initial guess (model without identification) and the true values (model with identification). The agreement between the numerical and theoretical computation is good in the whole frequency band. It is also clear that there is a difference between the residual force noise assuming the initial parameter estimates and the true ones (or the best-fit which completely overlaps it). The improvement in the estimation of the residual force noise is a factor $4$ around $0.4\,\unit{mHz}$ and a factor $2$ around $50\,\unit{mHz}$. The biggest contribution to this difference is due to the under-performing actuators at high frequency and the force coupling between the SC and the TMs at low frequency. The conclusion is that without any kind of system identification the residual force noise would be overestimated, especially at low frequencies.


\section{Concluding remarks}

This work has focused on the maximum likelihood parameter estimation in time domain
for the LPF mission. After introducing the dynamical equations and a model for
LPF, with its physical parameters and their significance, we have shown how to handle the effect of the controller, measure all known forces and subtract them from the data in order to provide an estimate of the residual force noise acting on the TMs. We have discussed our multi-experiment/multi-channel approach as a method to reach the desired measurement accuracy. We have started the discussion with a Monte
Carlo simulation of different noise realizations, showing the statistical consistency of the method. Considering
the physical situation where our knowledge of the system is not sufficient or the system is highly under-performing, we
have tested the algorithm with respect to the choice of the initial guess and we
have demonstrated that it has the ability to fully recover all parameters (within
one standard deviation) from a reasonable initial guess. Thereafter, to check
the robustness to non-gaussianities --- e.g., glitches in the interferometric
readout --- we have simulated corrupted time-series and repeated the analysis.
The result is that we are still able to recover the parameters and identify
those more sensitive to non-gaussianities. We have also proposed a method to handle
corrupted data that consists of a step-by-step reweighing estimation.The same methodology was employed to analyze data produced by a realistic LPF simulator in use at ESA: the scope is to enter into a mission-like routine, operating a system where many parameters are unknown or handled by a dedicated computer. Finally,
since the final scope of LPF is the characterization of the LISA Doppler link,
we have proven that the proposed parameter estimation is mandatory to correctly
assess the estimation of the differential residual force noise and avoid systematic errors at a level which would impact on the GW astronomy in the lower end of the LISA band.

\vspace{10pt}

\appendix


\section{LPF dynamics in detail} \label{app:dyn}

The controlled dynamics of LPF \cite{antonucci2011b,nofrarias2010} is described by the following equations
\begin{align}
& \mathbf{D}\cdot\mathbf{q} = \mathbf{g}~, \label{eq:eom_matrix1} \\
& \mathbf{g} = \mathbf{f}-\mathbf{C}\cdot(\mathbf{o}-\mathbf{T}\cdot\mathbf{o}_\text{i})~,\label{eq:eom_matrix2} \\
& \mathbf{o} = \mathbf{S}\cdot\mathbf{q}+\mathbf{o}_\text{n}~. \label{eq:eom_matrix3}
\end{align}
The total forces $\mathbf{g}$ produce the motion through the acting of the dynamics matrix $\mathbf{D}$ onto the physical coordinates $\mathbf{q}$. These forces are decoupled into external forces $\mathbf{f}$ (containing stochastic and deterministic signals) and control forces $\mathbf{C}\cdot(\mathbf{o}-\mathbf{T}\cdot\mathbf{o}_\text{i})$, where $\mathbf{C}$ is the control matrix and $\mathbf{o}_\text{i}$ are hardware injections at the level of the interferometer reference set-point, named controller \textit{guidance} inputs. $\mathbf{T}$ contains possible delays in the application of the thruster and the electrostatic suspension actuation. Finally, the interferometer readouts $\mathbf{o}$ are related to the physical coordinates $\mathbf{q}$ through the sensing matrix $\mathbf{S}$ and corrupted by the readout noise $\mathbf{o}_\text{n}$.

A straightforward procedure, described in \cite{monsky2009}, is capable of solving the problem of computing the out-of-loop external force per unit mass by subtracting the effect of the controller.


A mono-dimensional model for LPF along the sensitive axis was previously described \cite{nofrarias2010}. The dynamics is characterized by only two degrees of freedom
and the two interferometer readings are the $o_1$ channel (the relative
displacement of the optical bench to the reference TM) and the $o_{12}$ or differential channel (the relative displacement of the second TM to the reference TM). In the hypothesis of small motion,
the vectors and matrices can be written as
\begin{widetext}
\begin{align}
\mathbf{D} & =
\begin{pmatrix}
s^2+\left(1+\frac{m_1}{m_\text{SC}}+\frac{m_2}{m_\text{SC}}\right)\omega_1^2+\frac{m_2}{m_\text{SC}}\omega_{12}^2 & \frac{m_2}{m_\text{SC}}\left(\omega_1^2+\omega_{12}^2\right)+\Gamma_x \\
\omega_{12}^2 & s^2+\omega_1^2+\omega_{12}^2-2\Gamma_x
\end{pmatrix}~, \nonumber \\
\mathbf{f} & =
\begin{pmatrix}
f_1 - f_\text{SC} + \left(1+\frac{m_1}{m_\text{SC}}\right)f_{\text{SC}\rightarrow1} + \frac{m_2}{m_\text{SC}}f_{\text{SC}\rightarrow2} + f_{2\rightarrow1} \\
\minus f_1 + f_2 - f_{\text{SC}\rightarrow1} + f_{\text{SC}\rightarrow2} - \left(1+\frac{m_1}{m_2}\right)f_{2\rightarrow1}
\end{pmatrix}~, \nonumber \\
\mathbf{C} & =
\begin{pmatrix}
\minus\frac{A_\text{df}}{m_\text{SC}}\,C_\text{df}(s) & \frac{A_\text{sus}}{m_\text{SC}}\,C_\text{sus}(s) \\
0 & \frac{A_\text{sus}}{m_2}\,C_\text{sus}(s)
\end{pmatrix}~,\quad
\mathbf{T} =
\begin{pmatrix}
e^{\minus s\,\Delta t_1} & 0 \\
0 & e^{\minus s\,\Delta t_2}
\end{pmatrix}~,\quad
\mathbf{S} =
\begin{pmatrix}
1 & 0\\
S_{21} & 1
\end{pmatrix}~, \label{eq:eom_1D}
\end{align}
\end{widetext}
where $m_1\simeq m_2\simeq\unit[1.96]{kg}$ and $m_\text{SC}\simeq\unit[422.7]{kg}$ are the three body masses. $\omega_1^2$ and $\omega_2^2=\omega_1^2+\omega_{12}^2$ are parasitic \textit{stiffness} constants which model oscillator-like residual force couplings between each TM and the SC, mostly coming from gravitational, electrostatic and magnetic effects. $\Gamma_x\simeq\unit[4.9\e{\minus9}]{s^{\minus2}}$ models the gravitational coupling between the TMs. $f_1$, $f_2$, $f_\text{SC}$ are external forces on the first TM, the second TM and the SC; $f_{\text{SC}\rightarrow1}$, $f_{\text{SC}\rightarrow2}$ are coupling forces on the first and second TM by the SC; $f_{2\rightarrow1}$ is a residual coupling force between the two TMs. $C_\text{df}(s)$ and $C_\text{sus}(s)$ are the controller/actuator laws along the sensitive axis for the drag-free and suspension loops commanding the SC and second TM to follow the reference TM; $A_\text{df}$ and $A_\text{sus}$ are two gains for the application of the actuation of the thrusters and the electrostatic suspension forces. $\Delta t_1$ and $\Delta t_2$ are two delays in the previous actuation. $S_{21}$ is the lower off-diagonal element of $\mathbf{S}$ which models the sensing cross-talk from the $o_1$ to $o_{12}$ channel arising from imperfect common-mode rejection.


\section{Demonstration of noise non-stationarity} \label{app:demo}

We want to demonstrate that the variation of any of the parameters with respect to time produces non-stationary noise. Hence we check the validity of Eq.\;\eqref{eq:noise_variance} in Section \ref{sect:data_generation}. Expanding the noise around some nominal parameter value $p_0$ up to first order and by computing the variance of the noise, we get
\begin{align}\label{eq:appendix1}
\var[n] & \simeq \var\left[n_0\right]+\var\left[n'\delta p\right]+2\cov\left[n_0,n'\delta p\right] \nonumber \\
& = \var\left[n_0\right]+\var\left[n'\right]\delta p^2+2\cov\left[n_0,n'\right]\delta p~,
\end{align}
where $\var\left[n'\right]$ and $\cov\left[n_0,n'\right]$ are the variance of the noise first derivative and the covariance between the zero order and the first derivative. Now, for a zero-mean process with finite second moment, it holds
\begin{align}
\cov\left[n_0,n'\right] & = \text{E}\left[n_0 n'\right]-\text{E}\left[n_0\right]\text{E}\left[n'\right] \nonumber \\
& = \text{E}\left[\frac{1}{2}\frac{\partial}{\partial p} n^2\right] \nonumber \\
& = \frac{1}{2}\frac{\partial}{\partial p}\var[n]~.
\end{align}
Substituting this result back into Eq.\;\eqref{eq:appendix1}, we finally come to Eq.\;\eqref{eq:noise_variance}.


\bibliography{bibliography}

\end{document}